\documentstyle[12pt,epsf]{article}    
\newcommand{\ra}{\rangle}
\newcommand{\la}{\langle}

\newcommand{\II}{{\cal I}}

\newcommand{\HH}{{\cal H}}

\newcommand{\Id}{| {\cal I} \ra}

\newcommand{\be}{\begin{equation}}
\newcommand{\ee}{\end{equation}}
\newcommand{\ben}{\begin{eqnarray}\displaystyle}
\newcommand{\een}{\end{eqnarray}}
\newcommand{\refb}[1]{(\ref{#1})}
\newcommand{\p}{\partial}
\newcommand{\sectiono}[1]{\section{#1}\setcounter{equation}{0}}

\newcommand{\PSbox}[3]{\mbox{\rule{0in}{#3}\includegraphics{#1}\hspace{#2}}}

\begin{document}

{}~ \hfill\vbox{
\hbox{CTP-MIT-2993}}\break

\vskip 3.5cm
\centerline{\large \bf Tachyon Potentials, 
Star Products and Universality}

\vspace*{9.0ex}
\centerline{\large \rm Leonardo Rastelli and 
Barton Zwiebach\footnote{E-mail: rastelli@ctpivory.mit.edu,
zwiebach@mitlns.mit.edu}}

\vspace*{2.5ex}

\centerline{\large \it Center for Theoretical Physics}
\centerline{\large \it Massachussetts Institute of Technology}
\centerline{\large \it  Cambridge, MA 02139, USA}

\vspace*{10.5ex}
\medskip
\centerline {\bf Abstract}

\bigskip
\bigskip

We develop an efficient
recursive method to evaluate the tachyon
potential using the relevant universal 
subalgebra of the open string star algebra.
This method, using off-shell versions of Virasoro Ward identities,
avoids explicit computation of conformal transformations of operators
and  does not require a choice of background.
We illustrate the procedure with
a pedagogic computation of the level six tachyon
potential in an arbitrary gauge, and the evaluation 
of a few simple star products. We  give a background 
independent construction of the so-called identity 
of the star algebra, and show how it fits into family 
of string fields generating a commutative subalgebra.

\bigskip

\vfill \eject

\baselineskip=18pt

\tableofcontents  

\sectiono{Introduction} \label{s0}

In the last few months there has been a resurgence of interest
in string field theory. In particular, open string field theory
has provided a direct approach to study  
the physics of string theory tachyons. This includes
the case of tachyons living on the D-branes 
of bosonic string theory, and the case of tachyons living  on non-BPS 
D-branes or on  D-brane anti-D-brane pairs of superstring theories.

The old problem of the tachyonic instability
in bosonic open string theory has been put into a novel perspective
by Sen's conjecture that there is an extremum of the tachyon
potential at which the total negative potential energy
exactly cancels the tension of the D--brane~\cite{9902105}. Moreover,
solitonic lump solutions of the tachyon effective potentials
are identified with lower--dimensional branes~\cite{RECK,9902105}.
Similar conjectures exist for the tachyon living on
a coincident D-brane anti-D-brane pair, or on  non-BPS D-branes of
type IIA or IIB superstring
theories~\cite{9805019, 9805170,9808141,9810188,9812031,
9812135}.  String field theory has provided precise quantitative tests
of these conjectures \cite{9912249,0001201,0002237,
0002117,0003031,0001084,0002211,0003220,0004015,
0005036,chicagoBfield,wittenBfield}.  Indeed, the level expansion
scheme has turned open string field theory into a powerful
computational tool. This scheme is based on the `experimental'
realization that by truncating the string field to its low--lying modes
(keeping only the Fock states with $L_0 \leq l$), one obtains an
approximation that gets  more accurate as the level $l$ is
increased~\cite{KS, 9912249}.

For definiteness we focus here on the cubic open bosonic  
string field theory~\cite{WITTENBSFT}. In order to compute
an (off-shell) term coupling three space--time fields $\varphi_i$ in 
the string field action, one must evaluate the correlator of the
associated (typically non-primary) CFT operators ${\cal O}_i$ 
on a specific three punctured disk. Being 
cubic, the only additional  terms in the action are quadratic and
much easier to compute. 
In other string field theories there are higher order
terms, and our discussion will apply with minor modifications.

There are two main algorithms that have been applied for 
such computations. In one method,   
one first computes the (complicated) finite conformal 
transformations needed to insert with appropriate local coordinates the 
(non-primary) vertex operators ${\cal O}_i$,  and then evaluates the
correlators using the OPE's of various two dimensional fields.  
An alternative algorithm makes
use of an operator representation of the vertex as an
object $\langle V_3|$ 
in the 3--string (dual) Fock space~\cite{gross, cremmer, samuel, LPP}. 
The desired correlator
is given by the contraction  $\langle V_3|{\cal O}_1\rangle
\otimes |{\cal O}_2\rangle\otimes |{\cal O}_3\rangle$, 
where $|{\cal O}_i\rangle$
denotes the Fock space state associated to ${\cal O}_i$. This 
contraction requires purely
algebraic manipulations and  it can be naturally automated on a
computer \cite{0002237}.
The explicit expression for $\langle V_3|$, however, is tied to the
specific  choice of CFT background. 
This approach, therefore, does not implement the 
background independence feature of
tachyon condensation. 
The tachyon potential only involves string fields that
correspond to CFT states built by acting on the vacuum with the 
universal oscillators
$\{c_n, b_n\}$ and the (negatively moded)
background independent 
matter Virasoro operators $L_n^{matt}$ (with $c=26$)~\cite{9911116}. 
We denote the subspace generated by these as $\HH_{univ}$.
A main objective
of this paper is to provide a direct computational scheme using
$\HH_{univ}$.

\bigskip
Our procedure is the systematic off-shell implementation of the
conventional Virasoro Ward Identities that allow the
computation of correlators of Virasoro descendents in terms
of those of Virasoro primaries. We view the vertex $\langle V_3|$
as an object where a negatively moded Virasoro operator in one
state space can be
converted into linear combinations of positively moded 
Virasoro operators in all state spaces, with readily calculable
coefficients that capture the geometry of the
interaction. Such relations allow recursive computation
of all correlations involving the Virasoro operators. 
These relations are simply useful versions of the conservation laws
of the  operator formalism \cite{AlvarezGaume:1988bg,Vafa:1987ea}.  
They are obtained, for the Virasoro case, by studying contour integrals
of the type $\int T(z) v(z) dz$ where $v(z)$ is a globally defined vector
field on the punctured surface defining the interaction vertex. The 
identities arise by contour deformation and by referring
the objects inside the integrals to the coordinates chosen at
the punctures.  
Particular cases of the conservation laws
of the operator formalism have been used since very early 
times in string theory. For string interactions based on
contact type interactions (as  in light cone 
theories and classical closed string field theory) 
such relations have gone under the name of  `overlap conditions'.
By implementing analogous conservation laws for 
the ghost sector, we obtain a 
computational scheme totally within $\HH_{univ}$.
This makes the background independence of the tachyon
computations completely manifest. 
While the ideas behind our approach are certainly not new,
their applications are.

Computations using conservation laws are  elegant and simple
to carry out. As we shall illustrate,
the computations involved in \cite{9912249} become very
straightforward.
As opposed to the contraction method based on
the explicit Fock construction of $\langle V_3|$ the
method we discuss is naturally recursive, and
having done computatations to some level, only marginal additional
work is required to go one level higher.
In addition, compared with the set of all Lorentz scalars,
which is used in the best calculation to date of the
tachyon potential~\cite{0002237},
the $\HH_{univ}$ basis  becomes more and more economical as the level
is increased. We are thus led to believe the present method would allow
even higher level calculations.  We also hope that the  more 
transparent geometric understanding that is gained with 
conservation laws will help find an exact closed form expression
for the string field tachyon--condensate representing the stable
stationary point of the tachyon potential.

\medskip

With this goal in mind we begin some exploration of
the structure of star products using conservation laws.
The recognition \cite{9911116} of the important role
played by the background--independent subspace 
$\HH_{univ}^{(1)}$ of ghost number one states,
prompts some questions of a more
formal nature. The space $\HH_{univ}$,  built
just as $\HH_{univ}^{(1)}$ but containing states
of all ghost numbers, is readily identified as a natural
subalgebra of the full star algebra. Conservation laws
make this manifest. Since ghost number simply adds under
the star product,  $\HH_{univ}^{(0)}$ is  a subalgebra
of $\HH_{univ}$. In open string field theory
gauge parameters have ghost
number zero, therefore $\HH^{(0)}_{univ}$ is
a (the?) universal subalgebra of the open string gauge algebra.
Associated to once-punctured disks $\Sigma$ one has ghost number zero states,
usually referred to as 
`surface states', as they arise from Riemann surfaces. Since such states
$*$-multiply to give surface states, the set  $\HH (\Sigma)$
of such states is a subalgebra of $\HH^{(0)}_{univ}$.

It is natural to ask whether  the so--called 
`identity' element $\II$ of the string field $*$-algebra is  
 an element of $\HH_{univ}$. It
is. In fact $\II \in \HH(\Sigma)$. The identity is the state associated
to  a
unit disk with local coordinates that cover all of its interior.
 As such, it
can be written as an exponential of {\it total} (matter + ghost) Virasoro
generators acting on the vacuum. This description appears to be new. 
A denumerable basis for $\HH(\Sigma)$ is provided by $\HH^{(0)}(L) =\hbox{Span}[\{
L^{tot}\} |0\ra]$, the set of all total-Virasoro descendents of 
the SL(2,R) vacuum.  We point out that
the SL(2,R) vacuum and the identity $\II$  belong  to a family of `wedge--like'
surface states
$\HH_{wedge}$ of the CFT, each of which is associated to a 1--punctured
disk with
local coordinates  that cover a wedge of a certain angle within the unit circle.
The identity  corresponds to an angle of $360^\circ$ and the vacuum to $180^\circ$.
This family of wedge states is closed under $*$-multiplication and forms
a commutative algebra.
All in all we have the following inclusion of (universal) subalgebras of
the 
star algebra:
\be
\HH_{wedge} \subset \HH^{(0)}(L) \subset \HH_{univ}^{(0)} \subset
\HH_{univ}\,.
\ee

This paper is organized as follows.  In Section \ref{osfttachyon} 
we review various
descriptions of the three open string vertex, as well
as the universal description of the 
tachyon string field. As a comment on~\cite{9911116}, we 
explain that the
tachyon string field is spanned by the action of matter
and ghost Virasoro operators on the zero momentum tachyon.
In Section 3 we discuss in detail the derivation
of Virasoro conservation laws. We extend this
to  ghost fields and to
dimension one (non-primary) currents in Section 4.
As an illustration of our methods, in Section 5 we compute 
the open string tachyon potential to level (2,6). We do this
without gauge fixing. In Section 6
we discuss the identity element and
the subalgebra of wedge states. We
also  compute some  star products,
among them the  product of two zero-momentum tachyons.  We offer some
concluding remarks in Section 7.

\sectiono{Open string field theory and the tachyon}

\label{osfttachyon}

Section \ref{revisit} is  intended as
a  review of well--known results. 
In section \ref{background} we then present 
a new characterization of the `universal' subspace of states
$\HH_{univ}$
relevant for the tachyon condensation problem, as the states 
obtained acting with matter {\it and} ghost
Virasoro generators on the tachyon $c_1 | 0 \ra$.

\subsection{Open bosonic string field theory revisited} 

\label{revisit}

The dynamical variable of bosonic open string field theory (OSFT) is
the string field $|\Phi \ra$, which contains a component field 
for every state in the first--quantized string 
Fock space. The first--quantized Fock space of
the open string is just the state space $\HH$ of the combined
matter {\it and} ghost Conformal Field Theories (CFT's).
This state space $\HH$  can be broken up into subspaces of definite
ghost number. 
We will be  using conventions where the SL(2,R) 
vacuum $| 0 \ra$ carries ghost number zero,  the $b$ ghost
carries ghost number $-1$ and the $c$ ghost 
carries ghost number $+1$. With these conventions, a general
off--shell string field in OSFT 
corresponds to a state
in $\HH$ with ghost number $+1$.
A state in $\HH$ can be
represented as a local field acting on the vacuum
\be
|\Phi \ra = \Phi(0) | 0 \ra \, ,
\ee
where $\Phi(x)$ is defined on the boundary of the worldsheet.
We shall mainly be using conventions where the CFT is defined
on the upper--half complex plane, with the boundary of the
worldsheet mapped to real axis.\footnote{It is often useful
to think of states $ |\Phi \ra$ in terms of their
Schr\"{o}edinger representation, that is, as functionals
on the configuration space of strings.
Consider
the unit half--disk in the upper--half plane, 
$\{ |z| \leq 1, {\rm Im} \,z \geq
0 \}$, with the vertex operator  ${\Phi}(0) $
inserted at the origin. Impose standard open string boundary
conditions for the fields $\phi_i$ of the CFT on the real axis
($\phi_i$ is a collective label for all matter and ghost
fields), 
and impose some specific boundary conditions $\phi_b$ on the
outer boundary $|z| =1$.
 The path integral over $\phi_i$ on the interior of
 this half--disk with the boundary conditions
 $\phi_b$ held fixed produces some functional $\Psi_\Phi[\phi_b]$.
 This functional assigns a complex number to each
string configuration on the unit half--circle, and is therefore
the Schr\"{o}edinger representation of the
state $ \Phi(0) | 0 \ra $. }

The classical open string field theory action is a function
from  $\HH$ to the real numbers and is given by
\be \label{action}
S = -\frac{1}{g^2} \left( \frac{1}{2} \la \Phi, Q_B \Phi \ra +
\frac{1}{3} \,\la \Phi, \Phi, \Phi \ra \right)\,,
\ee
where $g$ is the open string coupling constant, $Q_B$ is the
BRST charge, and the 2-- and 3--point vertices
$\la \cdot, \cdot \ra$ and $\la \cdot, \cdot, \cdot \ra$ 
are defined in terms of CFT correlators. 

For the 2--point
vertex 
\be \label{kinetic}
\la \Phi , \Psi \ra \equiv \la I \circ \Phi(0)\,\, \Psi(0) \ra\,,
\ee
where $\la \;\ra$ on the right hand side
represents
the CFT correlator
and  $I$ denotes the SL(2,R) map $I(z) = -1/z$. 
The symbol
$f \circ \Phi(0)$, where $f$ is a complex
map, means the conformal transform
of  $\Phi(0)$ by $f$. For example if  $\Phi$ is a dimension $d$ primary
field, then 
$f \circ \Phi(0) = f'(0)^d \Phi(f(0))$. If  $\Phi$ is non--primary
the transformation rule will be more complicated and involve
extra terms with higher derivatives of $f$.
The cubic vertex is given by
\be \label{cubic}
\la \Phi_1 , \Phi_2 , \Phi_3\ra \equiv 
\la f_1 \circ \Phi_1(0) f_2 \circ \Phi_2(0) 
 f_3 \circ \Phi_3(0) \ra\,,
\ee
where $f_i$ are some specific conformal maps which
are described below (see (\ref{fW})). We shall also write
\be
\la \Phi_1 , \Phi_2 , \Phi_3\ra \equiv \la V_3 \; |\Phi_1\ra \otimes
|\Phi_2
\ra \otimes
|\Phi_3 \ra\,, 
\ee
where $\la V_3 |\in \HH^* \otimes \HH^* \otimes \HH^*$,
is a machine that given
three CFT states produces a real number. Another
familiar way of presenting the cubic vertex is in terms
of a  $*$-product defined as:
\be
\la \Phi , \Phi_1 * \Phi_2 \ra \equiv \la \Phi, \Phi_1, \Phi_2 \ra
\, , \quad \forall \Phi\,. 
\ee

\begin{figure} 
\begin{center} \PSbox{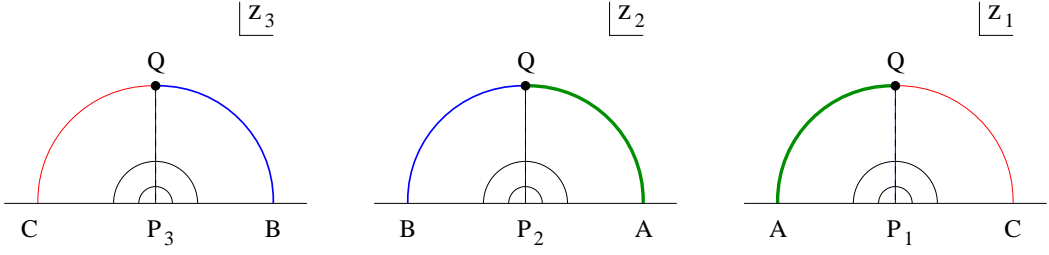 hscale=80
vscale=80}{5.5in}{1.1in} 
\vspace{0cm}
\end{center}
\caption{Representation of the cubic vertex as the gluing of
3 half--disks.} \label{pictorialfig}
\end{figure}

We can specify a three string vertex
with a picture showing how the worldsheets of
the three strings join together. 
For the SFT in hand \cite{WITTENBSFT}, the picture
is given in 
Fig.{\ref{pictorialfig}}.  The worldsheets of the three strings are
represented as unit half--disks
$\{ |z_i| \leq 1, \Im \, z \geq 0 \}$, $i=1,2,3$,
 in three copies of the complex plane. 
We glue the boundaries $|z_i| =1$
of  the three half-disks  with the identifications:
\ben 
\label{wittengluing}
z_1 z_2  =  -1\,, && \quad {\rm for} \; |z_1| =1, \, \Re \,z_1 \leq
0 \nonumber \\
z_2 z_3   =  -1\,,  && \quad {\rm for} \; |z_2| =1, \, \Re \,z_2 \leq
0 \\ z_3 z_1   =  -1\,,  && \quad {\rm for} \; |z_3| =1, \, \Re \,z_3
\leq 0 \nonumber
\een
Note that the common interaction point $Q$, 
defined by  $z_i=i$ (for
$i=1,2,3$) is the mid--point of each open string $|z_i| =1, \, \Im
\,z_i
\geq 0$.  The left half of the first string is glued with the 
right half of the second string, and the same is repeated cyclically.
This construction defines a specific `three--punctured disk', a
genus zero Riemann surface with a boundary, three marked points
(punctures) on this boundary, and a choice of local coordinates
$z_i$ around each puncture.

\begin{figure} 
\begin{center} \PSbox{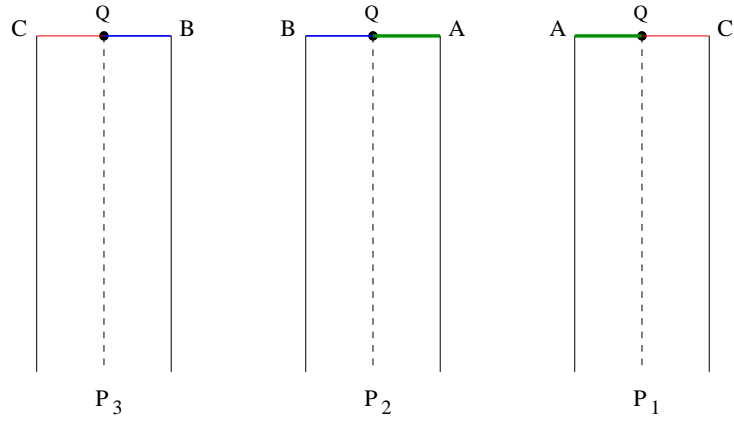 hscale=80 vscale=80}{4in}{2in} 
\vspace{0cm}
\end{center}
\caption{Representation of the cubic vertex as 
the gluing of 3 semi--infinite strips.} \label{strips}
\end{figure}

\begin{figure} 
\begin{center} \PSbox{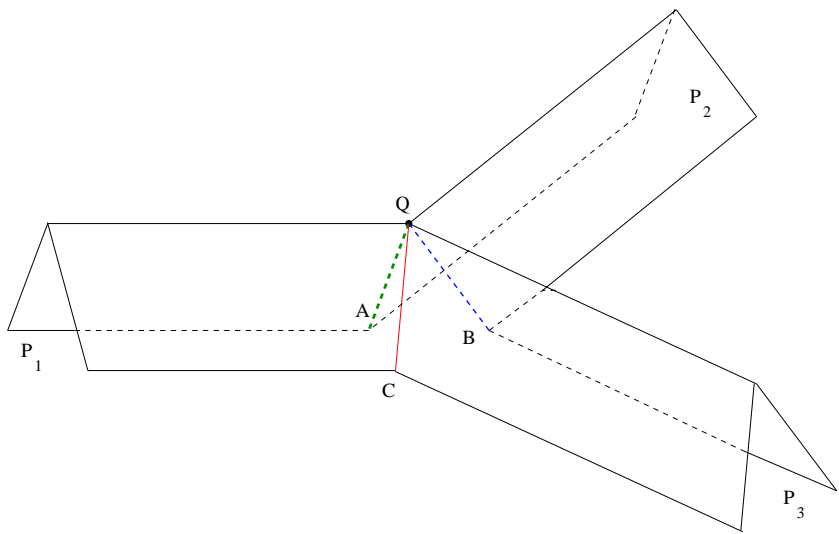 hscale=80 vscale=80}{4.5in}{2in} 
\vspace{0cm}
\end{center}
\caption{The result of gluing the 3 strips of Fig. \ref{strips}.}
\label{gluedstrips}
\end{figure}

It will be useful to recognize that on this glued
surface there is a globally defined Jenkins-Strebel
quadratic differential \cite{strebel}. This quadratic differential
$\varphi$ takes the form
\be
\label{js}
\varphi = \phi(z_i) dz_i^2 = - {1\over z_i^2} \, dz_i^2\,,
\ee
on each of the three coordinate patches. One readily verifies
that this assignment is consistent with the identifications
in \refb{wittengluing}. This quadratic differential has
second order poles at the punctures $z_i=0$. Its {\it horizontal
trajectories}, the lines along which $\varphi$ is
real and positive, foliate the surface and represent the open
strings. The quadratic differential has a first order zero at the
interaction point $Q$ 
($z_i=i$). Indeed, three neighborhoods of angle
$\pi$ are being glued at $Q$ 
and therefore a well-defined
coordinate
$w$ at $Q$ 
must be related to any given $z_i$ as $w \sim (z_i -
i)^{2/3}$. It then follows that  near $Q$ 
the quadratic
differential takes the form  $\varphi \sim 
dz_i^2
\sim w dw^2$, which shows the zero at $w=0$.
 A quadratic differential  defines a conformal
metric  $ds^2 =|\phi(z_i)| |dz|^2$. With this metric
the three half disks are presented as three semi-infinite strips
of width $\pi$, as in Fig. \ref{strips}.
 Gluing of these semi-infinite strips at the
edges produces the concrete (metric) representation of the
string vertex\footnote{This metric is actually a minimal
area metric, this fact is important to understand why the
Feynman rules of open string field theory generate a single
cover of the moduli spaces of Riemann surfaces with boundaries
\cite{gmw, Zwiebach:1991az}.}(Fig. \ref{gluedstrips}).

\medskip

\begin{figure} 
\begin{center} \PSbox{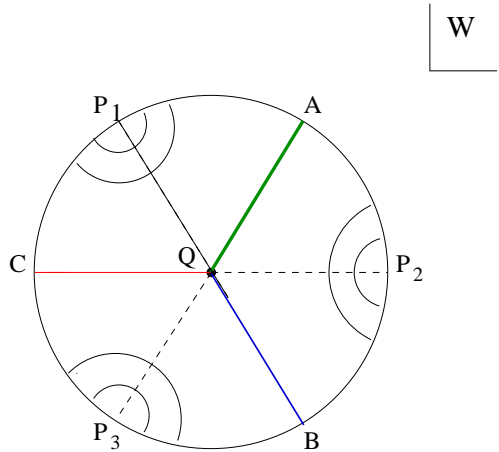 hscale=80 vscale=80}{2in}{2in} 
\vspace{0cm}
\end{center}
\caption{Representation of the cubic vertex as a 3--punctured
unit disk.} \label{3wedgesfig}
\end{figure}

Other conformal representations of the three string
vertex are useful. For example, we can  map the
three half disks to the interior of the unit disk
$|w| < 1$,
 as shown in Fig. \ref{3wedgesfig}.
Each worldsheet is sent to a $120^\circ$ wedge of this unit disk.
To construct the explicit maps that send $z_i$ to the $w$ plane,
one notices that the SL(2,C) transformation 
\be
\label{hdefi}
h(z)= \frac{1+iz}{1-iz}\,,
\ee
maps the unit upper--half disk $\{ |z| \leq 1, \Im z \geq 0 \}$ to the
`right' half--disk $\{ |w| \leq 1, \Re \, w\geq 0 \}$, with
$z=0$ going to $w=h(0) =1$. Thus the functions
\ben  
F_1^{120^\circ}(z_1) &  =&   e^{\frac{2 \pi i}{3}}\left(
\frac{1+iz_1}{1-iz_1} \right)^{\frac{2}{3}}\,, \nonumber \\
 F_2^{120^\circ}(z_2) &  =&  \left(
\frac{1+iz_2}{1-iz_2} \right)^{\frac{2}{3}}\,, \nonumber \\
 F_3^{120^\circ}(z_3) &  =& e^{-\frac{2 \pi i}{3}} \left(
\frac{1+iz_3}{1-iz_3} \right)^{\frac{2}{3}}\,, 
\een
will send the  three half-disks to three wedges in the $w$
plane of Fig. \ref{3wedgesfig}, with punctures at $e^{\frac{2 \pi
i}{3}}$, $1$, and $e^{-\frac{2 \pi i}{3}}$ respectively. 
Identifying the functions 
$f_i$ of (\ref{cubic}) as $f_i \equiv F_i^{120^\circ}$ we obtain
a definition of the cubic vertex.
In this representation cyclicity ({\it i.e.}, $\la \Phi_1, \Phi_2 ,
\Phi_3 \ra=
\la \Phi_2, \Phi_3 , \Phi_1 \ra$) is manifest by construction.
By SL(2,C) invariance, there are many other possible representations
that give exactly the same off--shell amplitudes.

\medskip

\begin{figure} 
\begin{center} \PSbox{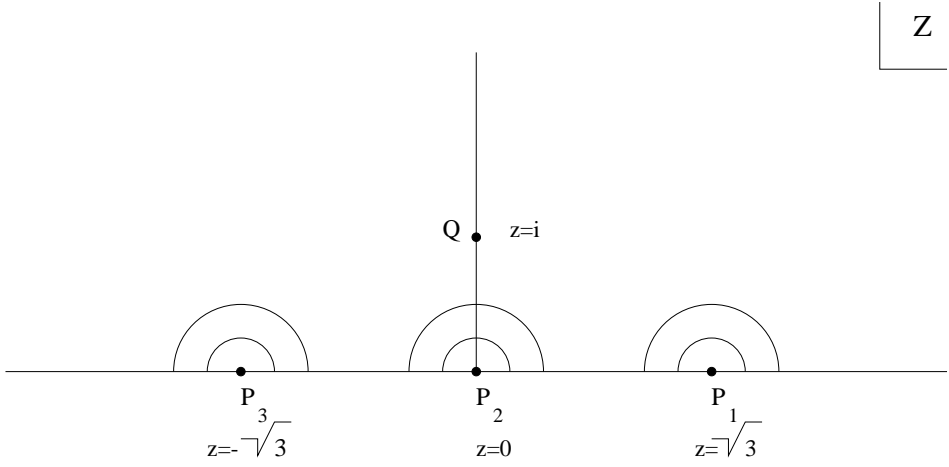 hscale=80 vscale=80}{5in}{2.2in} 
\vspace{0cm}
\end{center}
\caption{Representation of the cubic vertex as the upper--half
plane with 3 punctures on the real axis.}\label{uhpfig}
\end{figure}

A useful choice is to map the interacting $w$ disk
symmetrically to the upper half plane.  This is the convention
that we shall mostly be using.
We can therefore
define the functions 
$f_i$ 
by composing the earlier maps $F_i^{ 120^\circ}$ (that send the
half-disks to the $w$
unit disk) with the  map $h^{-1}(w) =-i
\,\frac{w-1}{w+1}$ taking this unit disk 
to the upper--half--plane, with the three punctures on the real
axis (Fig. \ref{uhpfig}),
\newpage
\ben 
\label{fW}
f_1 (z_1)& \equiv & h^{-1}\circ (F_1^{120^\circ}) (z_1)=
S (f_3 (z_1))  \cr\cr 
&=& \sqrt {3}+{\frac {8}{3}}\,z_1
+{\frac {16}{9}}\,\sqrt {3}\,{z_1}^{2}+{\frac
{248}{81}}\,{z_1}^{3}+{\frac {416}{243}}\,
\sqrt {3}\,{z_1}^{4}+{\frac {2168}{
729}}\,{z_1}^{5}+O\left ({z_1}^{6}\right )\,. \cr\cr
f_2 (z_2) & \equiv & h^{-1}\circ (F_2^{120^\circ})(z_2)  =S(f_1(z_2))= \tan
\left(\frac{2}{3} \, \arctan(z_2) \right)  \cr\cr
&=&  
{\frac {2}{3}}\, z_2-{\frac {10}{81}}\, {z_2}^{3}
+{\frac {38}{729}}\,{z_2}^{5}+O\left
({z_2}^{7}\right )\,.
 \cr\cr
f_3 (z_3) & \equiv & h^{-1}\circ (F_3^{120^\circ})\, (z_3) =
S (f_2 (z_3)) \cr\cr
&=&\hskip-9pt
-\sqrt {3}+{\frac {8}{3}}\, z_3-{\frac {16}{9}}\,\sqrt
{3}\, {z_3}^{2}+{\frac {248}{81}}\, {z_3}^{3}-{\frac {416}{243}}\,\sqrt
{3}\, {z_3}^{4} +{\frac {2168}{
729}}\, {z_3}^{5}+O({z_3}^{6})  \,.
\een
The three punctures are at 
 $ f_1(0)=+\sqrt{3}, f_2(0)=0, f_3(0)=-\sqrt{3},$ 
and the SL(2,R) map $S(z) = \frac{z - \sqrt{3}}{1+\sqrt{3}z}$
cycles them
(thus $S\circ S\circ S(z) =z$).  
This completes the definition
of the string field theory action.

\subsection{The universal tachyon string field}

\label{background}

The string field theory action (\ref{action}) 
describes the dynamics of
open strings in any conformal background.
In particular, we can take the CFT to be the Boundary Conformal
Field Theory (BCFT) of any bosonic D--$p$ brane, flat 26--dimensional
Minkowski space  being just the space filling D$25$ brane. Bosonic
D--$p$ branes are unstable  to decay and the open string tachyon  is
the signal of this instability.
The zero momentum tachyon state is $c_1 |0 \ra = c(0) |0\ra$.
This state belongs to a `universal' subspace $\HH_{univ}$ of
$\HH$ containing zero momentum scalars
\ben
 \HH_{univ} \equiv {\rm Span} \{ L^m_{-j_1} \dots L^m_{-j_p}\,
b_{-k_1} \dots b_{-k_q} \, c_{-l_1} \dots c_{-l_r} | 0 \ra\, , 
 \; j_i \geq 2, k_i \geq 2, l_i \geq -1 \} \,.
\een  
Note that the Virasoro generators belong to the matter CFT,
and the restriction $j_i \geq 2$ arises because $L_{-1}^m |0\ra=0$.
Similar restrictions apply to the ghost and antighost oscillators.
This is a `universal' subspace that is independent
of the details of the BCFT  (except for the central
charge $c=26$ of the matter stress--tensor). It is also
clear that for a string field in $\HH_{univ}$ 
the string field theory action (\ref{action}) takes
a background independent value, 
being determined by CFT correlators and operations
that involve only the ghosts and the matter stress--tensor.
This space $\HH_{univ}$ is that introduced by Sen in 
\cite{9911116}, with the distinction that we are now
including all ghost numbers. 

It is remarkable that  $\HH_{univ}$ {\it 
defines a  subalgebra of the star--algebra of open string fields}.
While this follows as a simple extension of the logic in 
\cite{9911116}, we will give in section 6 a direct argument 
based on explicit construction of products in the level expansion.
Note also that $Q_B : \HH_{univ} \to \HH_{univ}$, as the BRST charge
$Q_B$ is
built
from matter Virasoro and ghost oscillators. We can split 
$\HH_{univ}$ into a direct sum of spaces generated by 
states of a given ghost number:
\be
\HH_{univ} = \oplus_{n\in Z} \HH_{univ}^{(n)}\,.
\ee
Since ghost number adds under the star multiplication we
have that 
\be
\HH_{univ}^{(n)} \, * \, \HH_{univ}^{(m)} \subseteq  \,
\HH_{univ}^{(n+m)}\,, 
\ee
where it follows that $\HH_{univ}^{(0)}$ is a closed
subalgebra of $\HH_{univ}$. In section 6 we will discuss
natural subalgebras of $\HH_{univ}^{(0)}$.

One of the main results of \cite{9911116} is that
the classical tachyon string field, a field that 
must clearly be at zero momentum and have ghost number
one,  can be assumed
to lie on $\HH^{(1)}_{univ}$. All other fields can be 
set to zero consistently. 
We now claim that a basis for  $\HH^{(1)}_{univ}$ is given
by Fock states obtained acting on the tachyon $c_1 |0 \ra$
with all matter {\it and} ghost Virasoro generators,
\be \label{HunivVira}
 \HH^{(1)}_{univ} \equiv 
 {\rm Span} \{ L^m_{-j_1} \dots L^m_{-j_p}\,
 L^{gh}_{-l_1} \dots L^{gh}_{-l_p}\,
c_1 | 0 \ra \, \; j_i \geq 2, l_i \geq 1 \}\,.
\ee   
The significance of this statement is that all computations
of the tachyon potential reduce to computations of correlators
of stress tensors and operations using the
Virasoro algebra. The claim is equivalent to the statement that in
the pure ghost CFT, all the states of ghost number one
\be \label{H1ghost }
{\cal G}^{(1)} \equiv 
 {\rm Span} \{ 
b_{k_1} \dots b_{k_q} \, c_{l_1} \dots c_{l_r} | 0 \ra \;,
r-q=1 \} \,,
\ee
can be obtained as descendents of the tachyon $c_1 | 0 \ra$. Defining
\be \label{V1}
V^{(1)} \equiv {\rm Span} \{ L^{gh}_{-k_1} \dots L^{gh}_{-k_m} c_1 |0 \ra,
k_i \geq 1, k_1 \leq k_2 \dots \leq k_m \} \; ,
\ee
to be the Verma module in the ghost CFT built on the
primary $c_1 | 0 \ra$,  we are claiming that ${\cal G}^{(1)}= V^{(1)}$.
We can break up
both spaces into subspaces of definite $L_0$ eigenvalue,
which we denote by $V^{(1)}_n$  and  ${\cal G}^{(1)}_n$
for $L_0=n$. Clearly $V^{(1)}_n\subset {\cal G}_n^{(1)}$, so
it  suffices to show that  
${\rm dim}\, V^{(1)}_n = {\rm dim}\, {\cal G}_n^{(1)}$ for each $n$.
Let us first prove that the proposed basis states in (\ref{V1})
are linearly independent. 
If a linear combination  $|\psi \ra$ of 
descendents of $c_1 |0 \ra$
vanished identically,  $|\psi \ra$ would be a null state
of  the CFT (a trivial one), and it should show up in the Kac
determinant
formula.
 The roots of the Kac determinant are given by 
\begin{equation}
c = 1- \frac{6}{m(m+1)} \, , \quad
h_{r,s}(m) = \frac{[(m+1)r-ms]^2-1}{4m(m+1)}\,.
\end{equation}
For $c=-26$, $m=-\frac{1}{2} \pm \sqrt{\frac{17}{36}}$. Since
 $c_1  | 0 \rangle$ has $L_0=-1$, we are looking for integers $r,s$
such that $h_{r,s} = -1$, and  it is readily seen that there are no
solutions. 
Hence there are no null states in the Verma module of  $c_1 | 0
\rangle$, and the basis states in (\ref{V1}) are linearly
independent. It only remains to show that the dimensions
are the same\footnote{The small argument below is familiar from 
bosonization of the ghost coordinates.}.
 We can find the dimensions of $ V^{(1)}_n$
from the partition function
\begin{equation} \label{dimV}
\sum x^n {\rm dim}\, V^{(1)}_n = {1\over x}\prod_{k=1}^{\infty}
\frac{1}{1-x^k} \,.
\end{equation}
On the other hand we can determine the dimensions of
the subspaces ${\cal G}^{(k)}_{n}$ of ghost number $k$ and  $L_0=n$
by counting the ways we can act on the vacuum with  the oscillators
$b_{-j}$ and $c_{-m}$, where $j\geq 2$ and $m \geq -1$. Indeed
\be \label{Gpartition}
\sum_{n,k} x^n y^k {\rm dim}\,\,{\cal G}^{(k)}_{n} = 
\,\prod_{j=2}^{\infty}  
\prod_{m=-1}^{\infty} \left(1+ \frac{x^j}{y}\right)
 \left(1+ y \,x^m    \right) 
=\prod_{l=1}^{\infty}\frac{1}{1-x^l}\sum_{r =-\infty}^{+\infty}
 y^r x^{\frac{r^2-3r}{2}} \, ,
\ee
where in the second equality we have used  Jacobi triple
product identity (see {\it e.g.} (3.2.104) of \cite{gsw}).
We see that the term linear in $y$ gives
an expression for 
$\sum_n x^n \, {\rm dim}\,{\cal G}^{(1)}_{n}$ 
which precisely matches \refb{dimV}.
This concludes the argument
that all the states of ghost number one in the purely
ghost CFT can be identified with
the Verma module of $c_1 |0 \ra$\footnote{At ghost number 0 there
are states  in the ghost CFT that are {\it not} Virasoro descendents of the
vacuum $|0 \ra$ (the lowest dimensional primary at this ghost
number). The simplest example is $|\chi \ra = b_{-2}c_1 | 0 \ra = j^g(0)
| 0
\ra  $, where
$j^g$ is the ghost number current. This state is not a primary
since $j^g(z)$ is not a true tensor, indeed $L_1^{gh} |\chi \ra =|0 \ra$ .
However  $|\chi \ra $ is
 not
a descendent either, since $L^{gh}_{-1} |0 \ra =0$.
This phenomenon can occur in a non--unitary
CFT.}.

\sectiono{Virasoro conservation laws}

A straightforward procedure to evaluate the string field theory
interaction vertex (\ref{cubic}) is to compute
the finite conformal transforms of the three vertex operators
$\Phi_1$, $\Phi_2$, $\Phi_3$, and then evaluate the CFT correlator
using the relevant OPE's. Finite conformal transforms, however, are
complicated for non--primary operators and this method quickly
becomes very cumbersome at higher level. Conservation laws provide
an elegant and efficient alternative. Let us illustrate this for the
important case of generators of the Virasoro algebra, with central
charge $c$. We shall derive identities obeyed by the string vertex

$\la V_3 |$ of the 
general form
\be \label{generalcons}
\la V_3 | L^{(2)}_{-k} = \la V_3 | \, \left( A^{k} \cdot c + \sum_{n
\geq 0} 
a_n^{k} L^{(1)}_{n} + 
 \sum_{n \geq 0} b^{k}_n  L^{(2)}_{n} +  \sum_{n \geq 0} c^{k}_n
L^{(3)}_{n}
\right) \, ,
\ee
where $A^k$, $a^k_n$,  $b^k_n$ and $c^k_n$ are  coefficients
that will be determined below and depend on the geometry of the 
vertex.  (By cyclicity, the same identity
holds after letting $(1) \to (2)$,  $(2) \to (3)$, 
 $(3) \to (1)$). The point of this identity is that 
the negatively moded Virasoro generator $L^{(2)}_{-k}$
acting on state space 2 is
traded for a sum of positively moded generators acting on all the
three state spaces, plus a central term. 
Since all the states in the background--independent
subspace $\HH^{(1)}_{univ}$ are of the form (\ref{HunivVira}),
we see that by the conservation laws
for matter and ghost Virasoro generators, and
the commutation relations of the Virasoro algebra,
 we obtain a  recursive procedure
that allows one to express the coupling of any three states
in $\HH_{univ}^{(1)}$ in terms of the coupling $\la c_1, c_1, c_1
\ra$ of three tachyons.

\subsection{Setting up Virasoro conservations}

It is convenient to use the standard `doubling trick' for
open strings. We trade the holomorphic and
antiholomorphic components of the
stress tensor, defined in the upper--half
$z$ plane, for a single holomorphic field $T(z)$ defined
in the whole complex plane. With this convention,
the cubic vertex is regarded as a 3--punctured
sphere.
We examine a stress--tensor with general central term,
\be
T(z') T(z  ) \sim
\frac{c/ 2}{ (z'-z)^4} + \frac{2\, T(z) }{(z'-z)^{2}} +\frac{\partial
T(z)}{z'-z} +
\cdots
\ee
Under 
holomorphic change of variables,
\be
\tilde T(w) = \left( \frac{dz}{dw} \right)^2 T(z) +  {c\over 12} \,
S(z,w)\,,\qquad
S(z,w) =  { {dz\over dw}\, {d^3 z\over dw^3}  - {3\over 2} \bigl (
{d^2z\over
dw^2} \bigr)^2   \over \bigl( {dz\over dw} \bigr)^2 }\,.
\ee
The Schwartzian derivative $S(z,w)$ vanishes if $z$ and $w$ are related
by an SL(2,C) transformation. Under composition of conformal maps,
$z \to \rho(z)$, $\rho \to w(\rho)$, one finds
$S(w,z)  = \Bigl( {d\rho\over dz} \Bigr)^2 \, S(w,\rho) + S(\rho,z)$.

Consider the representation
(\ref{cubic}), (\ref{fW}) of the cubic vertex
in the full complex plane with punctures
at $+\sqrt{3}$, $0$ and $-\sqrt{3}$ (Fig. \ref{uhpfig}).
  We shall label
   the coordinate in the global plane as $z$, and the
local coordinates around the punctures as $z_i$, $1=1,2,3$.
 Let $v(z)$ be a holomorphic vector field $v(z)$, thus transforming as
$ \tilde v(w) = \left( \frac{dz}{dw} \right)^{-1} v(z) $.
We require $v(z)$ to be holomorphic everywhere
in the $z$ plane, except at the punctures where 
it may have poles. Since in our convention the punctures are all
located
at finite points on the real axis, we need to impose
regularity at infinity.
Performing  the change of variables $w=-1/z$, $\tilde v(w) = z^{-2}
v(z)$. Hence
for $v(z)$ to be regular at infinity, $\lim_{z \to \infty}
z^{-2} v(z)$ must be constant (or zero). 

The purpose of considering the vector $v(z)$ is that the 
product $v(z) T(z) dz$ 
transforms as a 1--form (except for a correction due to the central
term),
\be \label{Tvtransf}
 T(z) v(z) \, dz =  \Bigl( \tilde  T(w) - {c\over 12} \,
S(z,w)\Bigr) \tilde v(w) \, dw \, ,
\ee
and can be naturally integrated along 1--cycles. 
Moreover this 1--form is {\it conserved}, thanks to 
the holomorphy of $T(z)$ and $v(z)$, and integration
contours in the $z$ plane 
can be continuously deformed as long as we do not cross
a puncture. Consider
a contour ${\cal C}$ which
encircles the three punctures at $-\sqrt{3}$, $0$ and $\sqrt{3}$
in the $z$ plane of Fig. \ref{uhpfig}. For arbitrary vertex operators
$\Phi_i$, the correlator
\be \label{corr}
\la  \oint_{\cal C} v(z) T(z) dz \;
f_1 \circ \Phi_1(0) f_2 \circ \Phi_2(0) 
 f_3 \circ \Phi_3(0) \ra
\ee
vanishes identically, by 
shrinking the contour
${\cal C}$ to zero size around the point at
infinity (which is a regular point). 
In this argument  it is important
that  under the inversion $w=-1/z$, the Schwartzian
derivative vanishes and thus 
there is no contribution from the central term in (\ref{Tvtransf}).
Since the correlator (\ref{corr}) is zero for arbitrary
$\Phi_i$, we can write 
\be
\la  V_3 | \oint_{\cal C} v(z) T(z) dz =0 \,.
\ee
Deforming the contour ${\cal C}$ into the sum of three contours ${\cal
C}_i$
around the three punctures, and referring
the 1--form 
to the local coordinates, we obtain the 
basic relation
\be \label{basic}
\la  V_3 | \; \sum_{i=1}^3 \oint_{{\cal C}_i}   dz_i \; v^{(i)}(z_i)  
\, \Bigl( T(z_i) - {c\over 12} \,
S(f_i(z_i),z_i)\Bigr)   \; = 0 \;.
\ee
The maps $f_i$, since they differ
by SL(2,R) transformations, have the same Schwartzian derivative. 
We find\footnote{For this and most explicit
computations it is useful to use a symbolic manipulator
such as Maple or Mathematica.}
\be 
S(f_i,z_i)  = -\frac{10}{9} \frac{1}{(1+z_i^2)^2}
=-{10\over 9} + {20\over 9}z_i^2
-{10\over 3}z_i^4 + {40\over 9} z_i^6 + \cdots  \,, \quad  i=1,2,3.
\ee
Since this expression is regular at each puncture ($z_i=0$),
the central term can contribute 
to the conservation law (\ref{basic})
only for vector fields $v^{(i)}$ that have poles at the punctures.

We are looking for conservations laws of the form (\ref{generalcons}).
Recalling that 
\be
L_{-k}^{(i)}= \frac{1}{2 \pi i} \oint dz_i \, z_i^{-k+1} 
T^{(i)}(z_i) \, , 
\ee
we need a vector field which behaves as $v^{(2)} \sim z_2^{-k+1}
+O(z_2)$ around puncture 2, and has a zero in the other
two punctures, $v^{(1)} \sim O(z_1)$, $v^{(3)} \sim  O(z_3)$. 
A vector field of this type has (for $k>1$) a pole at the 
second puncture,
and is regular around the other two punctures. Contributions 
from the central term will then only appear in the second
state space.

\subsection{The first few conservation laws}

Consider using the globally defined vector field 
\be
v_1(z)
=-{2\over 9}\, (z^2 -3)\,.
\ee 
As discussed before, this has zero at punctures
1 and 3 and is regular at infinity.
Using the transformation law
$v_1^{(i)}(z_i) = v_1 (z(z_i))/ (dz/dz_i)$ and the 
relations \refb{fW}  
we derive the Taylor expansion of the
vector field referred to the each of the local
coordinates
\ben
v^{(1)}_1(z_1)& = & -\frac{4}{3\sqrt {3}} { z_1}+{\frac
{8}{27}}{{ z_1}}^{2}-{\frac {40}{81 \sqrt{3}}}\,{{ z_1}}^{3}+{\frac
{40}{729}}{{ z_1}}^{4}+{\frac {
104}{729 \sqrt{3}}}\,{{ z_1}}^{5}+O\left ({{ z_1}}^{6}\right )
\nonumber \\
v^{(2)}_1(z_2) & = & 1+{\frac {11}{27}}{{ z_2}}^{2}-{\frac {80}{729}}{{
z_2}}^{4}+{\frac {1136}{19683}}{{ z_2}}^{6}+O\left ({{ z_2}}^{8}\right
) \nonumber
\\
v^{(3)}_1(z_3)& = & \frac{4}{3\sqrt {3}} { z_3}+{\frac {8}{27}}{{
z_3}}^{2}+{\frac {40}{81 \sqrt{3}}}\,{{ z_3}}^{3}+{\frac {40}{729}}{{
z_3}}^{4}-{\frac {104
}{729 \sqrt{3}}}\,{{ z_3}}^{5}+O\left ({{ z_3}}^{6}\right ) 
\een
In this case the $v^{(i)}$ are regular around { \it each} puncture, 
so we get no contribution from the central term. Using
(\ref{basic}) and noting that 
integration amounts to the 
replacement $v_n^{(i)} z_i^n \to v_n^{(i)}
L_{n-1}^{(i)}$, we can immediately write the conservation law
 \ben
&&\, 0=  \la V_3|
 \Bigl( -{4\over 3\sqrt{3} }\,\,L_0 + {8\over 27} L_1 - {40\over
81\sqrt{3}} L_2 + {40\over 729}L_3 +  {104\over 729\sqrt{3}} L_4 \cdots
\Bigr)^{(1)} \nonumber \\
&&\quad+ \la V_3| \,
\Bigl( L_{-1} + {11\over 27} L_1 - {80\over 729} L_3 + {1136\over
19683}L_5 + \cdots \Bigr)^{(2)}\cr\cr
&&\quad+   \la V_3| \,  \Bigl(\,\,\, {4\over 3\sqrt{3} }\,\,L_0 +
{8\over 27} L_1 +
{40\over
81\sqrt{3}} L_2 + {40\over 729}L_3 -  {104\over 729\sqrt{3}} L_4 \cdots
\Bigr)^{(3)} \,.
\een
Thanks
to the  cyclicity of the string  vertex, analogous identities
hold by cycling the punctures, $(1) \to (2)$,  $(2) \to (3)$, 
 $(3) \to (1)$.
Using the vector field 
\be
v_2(z) = - {4\over 27} \,\, {z^2- 3 \over {z}}, 
\ee
we obtain 
\ben
&&\,0= \la V_3| \,\Bigl( -{8\over 27 }\,\,L_0 + {80\over 81\sqrt{3}}\,
L_1 -
{112\over
243} L_2 + {304\over 729\sqrt{3}}L_3 -  {400\over 19683} L_4 \cdots
\Bigr)^{(1)} \nonumber \\
&&\quad+ \la V_3| \,
\Bigl( L_{-2}\, + \,{5\over 54 }\, c \,+ {16\over 27}\, L_0 -
{19\over 243}
L_2 + {800\over 19683}L_4 + \cdots \Bigr)^{(2)}\cr\cr
&&\quad+ \la V_3| \,\Bigl( -{8\over 27 }\,\,L_0 - {80\over
81\sqrt{3}}\, L_1 -
{112\over
243} L_2 - {304\over 729\sqrt{3}}L_3 -  {400\over 19683} L_4 \cdots
\Bigr)^{(3)} \,.
\een
Since $v_2(z)$ has a pole at puncture 2 we got a contribution
from the central term.

In general we can get conservation laws for $L^{(2)}_{-k}$
with $v_k(z) \sim   (z^2-3)z^{-k+1}$.
For $k >2$, using this vector in (\ref{basic})
one obtains an identity that besides  $L^{(2)}_{-k}$ involves 
other negatively moded Virasoro generators $L^{(2)}_{-k+2}$,
 $L^{(2)}_{-k+4}$, $\dots$. It is straightforward to remove these terms
by subtracting the conservation laws for smaller $k$. 
Indeed, using the vectors
\ben
v_3(z) &=& -{8\over 81}\, {z^2-3\over z^2} - {7\over 9}\,\, v_1(z)\,,  \cr\cr
v_4(z) & =& -{16\over 243} \, {z^2-3\over z^3} - {26\over 27}\,\, v_2(z)
\,, 
\een
we obtain the conservation laws:
\ben
&&\, 0= \la V_3 | \,
\Bigl( \,\,\,{68\over 81\sqrt{3} }\,\,L_0 + {40\over 243}\, L_1
-{152\over 243\sqrt{3}} L_2 + {8792\over 19683}L_3 -  {3320\over
6561\sqrt{3}}
L_4 \cdots \Bigr)^{(1)} \nonumber \\
&&\quad + \la V_3 | \, \Bigl( L_{-3} - {80\over 243} L_1 + {2099\over
19683} L_3 -
{3568\over
59049}L_5 + \cdots \Bigr)^{(2)}\cr\cr
&&\quad+ \la V_3 | \,
\Bigl( -{68\over 81\sqrt{3} }\,\,L_0 + {40\over 243}\, L_1 +
{152\over 243\sqrt{3}} L_2 + {8792\over 19683}L_3 +  {3320\over
6561\sqrt{3}}
L_4 \cdots \Bigr)^{(3)} \,.
\een

\ben
&&\, 0= \la V_3 | \,\Bigl(  {176\over 729 }\,\,L_0 - {416\over
729\sqrt{3}}\, L_1 -
{800\over 19683} L_2 + {13280\over 19683\sqrt{3}}L_3 -  {84448\over
177147} L_4 \cdots \Bigr)^{(1)} \,  \cr \cr
&& \quad + \la V_3 | \,\Bigl( L_{-4} \,-\,{5\over 27}\, c - {352\over
729} L_0 + {1600\over 19683} 
L_2 - {8251\over 177147}L_4 + \cdots \Bigr)^{(2)}\cr\cr
&&\quad +  \la V_3 | \,\Bigl( {176\over 729 }\,\,L_0 + {416\over
729\sqrt{3}}\, L_1 -
{800\over 19683} L_2 - {13280\over 19683\sqrt{3}}L_3 -  {84448\over
177147} L_4 \cdots \Bigr)^{(3)}
\een
Higher identities can be derived analogously. The identities above
suffice for a level (4,8) computation of the string action.

\medskip

We conclude this section with a discussion of the so--called
`reparametrization invariances' of the cubic
vertex \cite{wittensupersft}.  It is
well known that for $c=0$ ({\it total} Virasoro generators)
the combination
\be
K_n = L_n -(-1)^n L_{-n}\,, 
\ee
is conserved on the vertex, that is
\be \label{Kn}
\la V_3 | \left( K_n^{(1)} + K_n^{(2)}+  K_n^{(3)}\right)
 =0 \,.
\ee 
These relations are special cases of the Virasoro conservation laws
and can be obtained by adding the 3 cyclic versions of the
$L_{-n}$ conservation. A direct and more elegant derivation
is as follows.
Consider the vector field defined by $v_n^{(i)}(z_i) = z_i^{n+1} -(-1)^n
z_i^{-n+1}$ around each of the punctures.
This vector field is globally defined since the expressions
on each puncture are consistent with the gluing relations
(\ref{wittengluing}) of the string vertex.
We have then
\be
\la V_3 | \; \sum_{i=1}^3 \oint T^{(i)}(z_i) \, \Bigl( z_i ^{n+1} -
(-1)^n z_i^{-n+1} \Bigr) =0 \, ,
\ee
which immediately gives (\ref{Kn}).\footnote{Since the vector field 
$v_n^{(i)}$ was
directly given in terms of the local coordinates $z_i$ and shown to be
globally defined the reader may wonder how the central term
violations \cite{gross} of these identities would arise.
Indeed, while the
contour integrals can be canceled pairwise at the boundary
of the local disks without extra contributions (the transition functions are
projective),  
there is a subtlety at the interaction points (the points
$z_i=\pm i$ on the local coordinates and $z=0, \infty$ on the
global disk). To deal with this properly one can
cancel the contour integrals pairwise, but not all the
way to the interaction points.  This leaves three
tiny contour  integrals $\sum_i \int T(z_i) v(z_i) dz_i$
that add up to a contour surrounding each
interaction point.  To evaluate this one has to pass
again to the coordinate $z$ vanishing at the interaction point.
A simple computation shows that the Schwarzian
$S(z_i, z)$ has a second order pole at $z=0$. In addition, the
vector $v$ has a  first order zero at $z=0$. Thus a central
charge contribution (in fact, of the right value) arises.}

\sectiono{Ghost and current conservation laws}

The $b$ ghost field is a true conformal tensor of dimension two.
Thus its conservation laws are identical to those for the 
$c=0$ Virasoro generators, with the formal replacement
$L_k^{(i)} \to b^{(i)}_k$. In this section we derive conservation
laws for the $c(z)$ field. In addition we derive conservation
laws for currents.

\subsection{Conservations for the $c$-ghost}

 The $c$ ghost is a primary field of dimension minus one,
$\tilde c (w) = \left( \frac{dz}{dw} \right)^{-1} c(z) \,.$
To derive conservation laws, we consider a globally defined
`quadratic differential' on the sphere
\be
\varphi = \phi(z) (dz)^2 =
\phi'(z') (dz')^2 \, ,
\ee
holomorphic everywhere except for possible poles at the
punctures. Regularity at infinity requires the $\lim_{z \to \infty}
z^4 \phi(z)  $ to be finite.
The product $c(z)  \phi(z) dz$ is a 1--form, 
which is conserved thanks to holomorphy of $c(z)$ and $\phi(z)$.
We can then use contour deformations and
following exactly
the same logic as for the 1--form $v(z) T(z) dz$ 
considered in the previous section, 
we obtain 
\be \label{basicc}
\la V_3 | \, \sum_{i=1}^{3} \, \oint_{{\cal C}_i}
 dz_i \, c^{(i)}(z_i) \phi^{(i)}(z_i)
=0 \, .
\ee

For example, with the quadratic 
differential 
\be
\phi_0(z) =-3 \,  z^{-2} (z^2 - 3)^{-1}
\ee 
one obtains:
\ben \label{firstc}
&&\, 0=
\la V_3| \Bigl( -{4\over 3\sqrt{3} }\,\,c_1 + {8\over 27}\, c_2 +
{68\over 81\sqrt{3}}\, c_3 - {176\over 729}\, c_4 +\cdots  \Bigr)^{(1)}
\cr \cr
&&\quad +\la V_3 |
\Bigl( c_0 \,-\,{16\over 27}\, c_2 + {352\over 729} c_4 -
{8368\over 19683} c_6 +  \cdots
\Bigr)^{(2)}\cr\cr 
&&\quad + \la V_3 | \Bigl( {4\over 3\sqrt{3} }\,\,c_1 + {8\over 27}\,
c_2 -
{68\over 81\sqrt{3}}\, c_3 - {176\over 729}\, c_4 +\cdots \Bigr)^{(3)}
\,.
\een

Higher conservation laws are obtained with quadratic differentials
having higher order poles at $z=0$. With
\ben
\phi_1(z) &=& -2 \,  z^{-3} (z^2 - 3)^{-1}\,, 
\cr\cr
\phi_2(z) & =& - {4\over 3} \,\,  z^{-4} (z^2 - 3)^{-1} + {2\over 9} \,
\phi_0 (z) \,, 
\een
we derive
\ben \label{c-1}
&&\,0= \la V_3| \,\Bigl( -{8\over 27 }\,\,c_1 + {80\over
81\sqrt{3}}\, c_2 -
{40\over 243}\, c_3 - {416\over 729\sqrt{3}}\, c_4 +\cdots
\Bigr)^{(1)}
\cr \cr
&&\quad +\la V_3| \,\Bigl( c_{-1} \,-\,{11\over 27}\, c_1 + {80\over
243} \,c_3
-{5680\over 19683}\, c_5  + \cdots
\Bigr)^{(2)}\cr\cr
&&\quad + \la V_3| \, \Bigl( -{8\over 27 }\,\,c_1 - {80\over
81\sqrt{3}}\, c_2 -
{40\over 243}\, c_3 + {416\over 729\sqrt{3}}\, c_4 +\cdots \Bigr)^{(3)}
\,.
\een 
\ben
&&\, 0= \la V_3 | \, \Bigl( -{40\over 81\sqrt{3} }\,\,c_1 + {112\over
243}\, c_2 -
{152\over 243\sqrt{3}}\, c_3 + {800\over 19683}\, c_4 +\cdots
\Bigr)^{(1)}
\cr\cr
&&\quad +\la V_3 | \,\Bigl( c_{-2} \,+\,{19\over 243}\, c_2 -
{1600\over 19683}
\,c_4
+{4640\over 59049}\, c_6  + \cdots
\Bigr)^{(2)}\cr\cr
&&\quad + \la V_3 | \, \Bigl( {40\over 81\sqrt{3} }\,\,c_1 + {112\over
243}\, c_2 +
{152\over 243\sqrt{3}}\, c_3 + {800\over 19683}\, c_4 +\cdots
\Bigr)^{(3)} \,.
\een

\medskip

The linear combinations 
\be \label{Cndef}
C_n \equiv c_{n} + (-1)^n c_{-n} \;,
\ee
are analogous to the $K_n$'s discussed at the end of the previous
section, since they are conserved on the three string vertex,
\be
\label{Cn}
\la V_3 | \left( C_n^{(1)} + C_n^{(2)}+  C_n^{(3)}
\right)  =0 \,.
\ee 
This identity is easily derived considering the quadratic
differential defined by the {\it same} functional form
\be \label{Cndiff}
\phi^{(i)}_n( z_i) = z_i^{-2+n} + (-1)^n z_i^{-2-n} \quad i=1,2,3
\ee
in the three local coordinate patches. 
These expressions are consistent with the gluing conditions 
(\ref{wittengluing}). Thus this quadratic
differential is globally defined and 
application of the basic relation (\ref{basicc}) immediately
gives (\ref{Cn}). 
In particular for $n=0$, $\phi^{(i)}_0$ is the canonical quadratic
differential \refb{js} on the string vertex, which gives
the conservation law
\be
\label{c0}
\la V_3 | \left( c_0^{(1)} + c_0^{(2)}+  c_0^{(3)}
\right)  =0 \,.
\ee

\subsection{Current conservation}

Let $j(z)$ be a  holomorphic dimension one current.  
Moreover, assume that $j(z)$  is not primary and take the OPE
of the stress tensor with the current to be:
\be 
T(z')\, j (z) = \,{2q\over (z'-z)^3}\, + \,  {j (z) \over (z'-z)^2 }
+\, { \p\, j (z)\over (z'-z)}\, + \cdots \, ,  
\ee
where $q$ is a dimensionless number. For example,
$q= -3/2$ for the ghost number current of the $(b,c)$ system. The finite
transformation associated to $j(z)$ is
\be
{dz\over dw} \cdot
 j (z) = \tilde \jmath (w) - q \,\, {d^2z\over dw^2} 
\Bigl( {dz\over dw}\Bigr)^{-1}\, .
\ee
If $f(z)$ is an analytic {\it scalar} ($f(z) = \tilde
f(w)$),  than $j(z) f(z) dz$ transforms as
a 1--form (except for a   correction due
to the anomaly $q$). We assume that $f(z)$ is
holomorphic everywhere (including infinity) except for
possible poles at the punctures and follow
the usual strategy to derive conservation laws.
Consider again a contour ${\cal C}$ which encircles
counterclockwise the 3 punctures in the $z$ plane of Fig. \ref{uhpfig}. 
This time $\la V_3 |\oint_{{\cal C}}  dz\,  j(z) f(z) $
is not identically zero, since to shrink the contour
around the point at infinity we need to perform
the transformation $z \to \tilde z = 1/z$,
and the current is not covariant under this. One finds instead
\be 
\la V_3 | \left(
\oint_{{\cal C}}  dz\,  j(z) f(z) + 2 q \, 
\int_{\tilde z=0} d \tilde z \;  {
\tilde f(\tilde z) 
\over \tilde z} \right) = 0
\ee
where $\tilde z = 1/z$. Here the second integral 
is evaluated along a contour that goes around
the $\tilde z$ plane origin counterclockwise.
This term will matter only when $\tilde f(\tilde z)$ 
is regular or has poles at 
$\tilde z = 0$, in other words when $f(z)$ is regular or worse
at infinity. We can now deform the contour ${\cal C}$ into
the sum of 3 contours ${\cal C}_i$ around the 3 punctures
and pass to the local coordinates. The conservation law reads
\be
\label{jcon}
\la V_3 | \, \left( 2q \int_{\tilde z=0} d\tilde z\; 
 { \tilde f(\tilde z) 
\over \tilde
z}
+\sum_{i=1}^3\int  \,\Bigl[ j(z_i)  - q \,\, {d^2z\over dz_i^2} 
\Bigl( {dz\over dz_i}\Bigr)^{-1} \,\Bigr] \, f^{(i)}(z_i) 
dz_i  
\right) =0 \,.
\ee
Note that there are two contributions to the anomaly.

\medskip
\noindent

The simplest conservation law arises  for the scalar
function $f_0(z) =1$. The anomaly contribution arises from
the first term in the above equation. We find  
\be
0=\langle V_3| \Bigl(  j_0^{(1)}+j_0^{(2)}+j_0^{(3)} + 2q \Bigr)\,, 
\ee
which reflects the familiar anomaly in charge conservation.
\bigskip
With the following functions 
\ben
f_1(z) &=&  {2\over 3}\,\cdot\,  {1\over z}\, , \cr\cr
f_2 (z) &=& - {4\over 27} \, {z^2-3\over z^2} \,,\cr\cr
f_3(z) &=& -{8\over 81} \, {z^2-3\over z^3}\,
 - \, {11\over 27} \, f_1(z)\,,\cr\cr
f_4(z) &=& -{16\over 243} \, \, {z^2-3\over z^4}
 -\, {16\over 27}\, f_2(z) \,,
\een
we obtain the conservation laws:
\newpage
\ben
&& \, 0=\la V_3 |\Bigl(  {2\over 3\sqrt{3} }\,\,\,j_0 - {16\over 27}\,
\,j_1 +
{32\over 81\sqrt{3}}\, \,j_2 + {16\over 729}\, \,j_3 - {64\over
729\sqrt{3}}\,
\,j_4+\cdots   \Bigr)^{(1)}
\cr\cr
&& \quad +\la V_3 |\Bigl( \,j_{-1} \,+\,{5\over 27}\, \,j_1 - {32\over
729} \,j_3 +
{416\over 19683} \,j_5 +  \cdots
\Bigr)^{(2)}\cr\cr 
&& \quad + \la V_3 | \Bigl( -{2\over 3\sqrt{3} }\,\,\,j_0 - {16\over
27}\, \,j_1 -
{32\over 81\sqrt{3}}\, \,j_2 + {16\over 729}\, \,j_3 + {64\over
729\sqrt{3}}\,
\,j_4+\cdots \Bigr)^{(3)} \,.
\een
\ben \label{j-2}
&& \,0= \la V_3 |\Bigl(  -{64\over 81\sqrt{3} }\,\,\,j_1 + {128\over
243}\, \,j_2 -
{320\over 729\sqrt{3}}\, \,j_3 - {256\over 19683}\, \,j_4\,+\cdots  
\Bigr)^{(1)}
\cr\cr
&& \quad +\la V_3 |\Bigl( \,j_{-2} \,+ {22\over 27} q\,+\,{2\over 9}\,
\,j_0 -
{13\over 243} \,j_2 + {512\over 19683} \,j_4 +  \cdots
\Bigr)^{(2)}\cr\cr 
&& \quad + \la V_3 |\Bigl( {64\over 81\sqrt{3} }\,\,\,j_1 + {128\over
243}\, \,j_2 +
{320\over 729\sqrt{3}}\, \,j_3 - {256\over 19683}\, \,j_4 +\cdots
\Bigr)^{(3)}\,.
\een
\ben
&& \,0 = \la V_3 |\Bigl(  -{22\over 81\sqrt{3} }\,\,\,j_0 + {16\over
243}\, \,j_1 +
{160\over 243\sqrt{3}}\, \,j_2 - {10288\over 19683}\, \,j_3
+{3136\over 6561\sqrt{3}}\, \,j_4\,+\cdots  
\Bigr)^{(1)}
\cr\cr
&& \quad +\la V_3 |\Bigl( \,j_{-3} \,- {32\over 243}  \,j_1 + {893\over
19683} \,j_3 - {1504\over 59049} \,j_5 +  \cdots
\Bigr)^{(2)}\cr\cr 
&& \quad + \la V_3 |\Bigl( {22\over 81\sqrt{3} }\,\,\,j_0 + {16\over
243}\, \,j_1 -
{160\over 243\sqrt{3}}\, \,j_2 - {10288\over 19683}\, \,j_3
-{3136\over 6561\sqrt{3}}\, \,j_4 +\cdots \Bigr)^{(3)} \,.
\een
\ben
&& \, 0 =\la V_3 |\Bigl(  {256\over 729\sqrt{3} }\,\,\,j_1 - {512\over
19683}\, \,j_2 -
{12544\over 19683\sqrt{3}}\, \,j_3 + {91136\over 177147}\,
\,j_4\,+\cdots  
\Bigr)^{(1)}
\cr\cr
&& \quad +\la V_3 |\Bigl( \,j_{-4} - {562\over 729} q \,- {38\over 243}
\,j_0 +
{1024\over 19683} \,j_2 - {5125\over 177147} \,j_4 +  \cdots
\Bigr)^{(2)}\cr\cr 
&& \quad + \la V_3 |\Bigl( -{256\over 729\sqrt{3} }\,\,\,j_1 - 
{512\over 19683}\, \,j_2 +
{12544\over 19683\sqrt{3}}\, \,j_3 + {91136\over 177147}\, \,j_4
 +\cdots \Bigr)^{(3)} \,.
\een
Both for $j_{-2}$ and $j_{-4}$ the anomaly receives contributions
from the two terms in \refb{jcon}.

\newpage
\sectiono{Sample computation: Level six tachyon potential} 

For the purposes of illustration we will
compute here the open string field action relevant
to tachyon condensation to level $(2,6)$. 
In order to provide not only
an illustration but also new information, we will
compute the action without imposing a gauge choice.

\subsection{Notation and the basic three point function} 

We begin with some preliminaries.  Given three vertex operators
$\widehat A, \widehat B, \widehat C$  we denote the corresponding
states by $A|0\rangle, B|0\rangle,\, C|0\rangle $, where $A,B,C$
denote  sets of oscillators.  As explained before (see section 2.1), 
\ben
\langle \widehat A, \widehat B, \widehat C \rangle
&=& \langle V_3 |\,\,\,  A^{(1)}|0\rangle_{(1)}\otimes
B^{(2)}|0\rangle_{(2)}\otimes\,  C^{(3)}|0\rangle_{(3)} \,,\cr\cr
&=& \langle V_3 |\,\,\,  A^{(1)}B^{(2)}C^{(3)}\,\,
|0\rangle_{(3)}\otimes
|0\rangle_{(2)}\otimes\,  |0\rangle_{(1)}\,,
\een
where in the second step we moved the vacua all the 
way to the right.\footnote{In the operator
formulation of bosonic open string field theory it is natural to use a
Grassmann
even string field. In this case, the in-vacuum $|0\rangle$ is Grassmann
odd. The
rearrangement  of the vacua in
the second equation is convenient to avoid explicit signs.}
For Grassmann odd operators (the 
operators relevant to the classical action) the following
cyclicity and twist relations hold
\be
\langle \widehat A, \widehat B, \widehat C \rangle  = 
\langle \widehat B, \widehat C, \widehat A \rangle \,, \quad
\langle \widehat A, \widehat B, \widehat C \rangle = 
\Omega_A\Omega_B\Omega_C\, 
\langle \widehat C, \widehat
B, \widehat A \rangle\,, 
\ee
where $\Omega$ is the twist eigenvalue and equals $(-)^{N +1}$,
where $N$ is the total number operator measured with respect
to the SL(2,R) vacuum. 
Also with a little abuse of notation we will denote
$\langle \widehat A, \widehat B, \widehat C \rangle$
simply by $\langle  A,  B,  C \rangle $, where we simply
write the Fock space oscillators instead of the operators.

The basic correlator that we will need is
\be
\label{bcor}
\langle c_1, c_1, c_{1} \rangle = \langle V_3| \,\, 
c_1^{(1)} \, c_1^{(2)} \,  c_{1}^{(3)}\,\,
|0\rangle_{(3)}\otimes |0\rangle_{(2)} \otimes|0\rangle_{(1)}  
\ee
For this correlator we use the conformal field theory
definition and thus write it as:
\begin{eqnarray}
\langle c_1, c_1, c_{1} \rangle &=& \langle f_1\circ c (0), f_2\circ c
(0),  f_3\circ c (0) \rangle \cr\cr
&=& \Bigl\langle\,\, { c (\sqrt{3})\over {8 \over  3}}\,\,, { c (0)\over
{2 \over 3}}, { c (-\sqrt{3})\over {8 \over 3}} \Bigr\rangle \cr
& =& {3^3 \over 2^7} 
\langle c(\sqrt{3}) c(0) c(-\sqrt{3}) \rangle = 
{3^4 \sqrt{3} \over 2^6}\,.
\end{eqnarray}  
In obtaining this we used 
\be 
f_i\circ c(0) = c(f_i(0)) /f'_i(0), \quad 
\langle c(z_1) c(z_2) c(z_3)\rangle = (z_1-z_2)(z_1-z_3)
(z_2-z_3)\,.
\ee
and equations (\ref{fW}) to read the values of the $f_i'(0)$. 
The answer is
written as
\be
\langle c_1, c_1, c_{1} \rangle = K^3\,, \quad  K \equiv
{3\sqrt{3}\over 4}\,.
\ee
The conservation laws allow the computation of all necessary
three point functions in terms of this single one. No other 
three point function need to be evaluated directly for 
the problem of finding the tachyon potential.

\subsection{Level (2,6) tachyon potential in arbitrary gauge}

The relevant level two string field\footnote{We have changed the
normalization
of the $v$ field by ommitting a $\sqrt{13}$ that was included in
\cite{9912249}.}
is then:
\ben \label{e21}
|T\rangle = t c_1 |0\rangle + u c_{-1}|0\rangle + v 
 L_{-2}^m\, c_1\,|0\rangle  + w \, c_0 b_{-2} c_1 |0\rangle . 
\een
Note that $w$ is the only field that would not
have appeared in the Siegel gauge. We will now compute the
string field theory action for such a string field. Even though
the action is not gauge fixed, gauge invariance is actually 
broken by the level truncation.
The tachyon potential ${\cal V}(T)$  we will compute is given by
\be
{{\cal V}(T)\over 2 \pi^2 M} \equiv {1\over 2\pi^2} f(T) =V(T)=
\,{1\over 2}
\langle T, Q_B T\rangle + {1\over 3}\, \langle\, T, T * T \rangle \,,
\ee
where $M$ denotes the brane mass.

The computation of the kinetic terms requires the
BRST operator. To the required level, and at zero
momentum this is given by
\begin{eqnarray}
 Q_B &  =& c_0\, 
 L_0^{tot}    -2\,b_0 c_{-1} c_1 -4\,b_0 c_{-2} c_2\cr
&&\quad -3\, b_{-1}c_{-1} c_2  - 3 c_{-2} c_1
b_1 + L_{-2}^m \,c_2 + \,c_{-2} 
L_2^m \,.   
\end{eqnarray}
In addition, the rules of BPZ conjugation for mutually (anti)commuting
oscillators is simply to let $\phi_n \to (-)^{n+h} \phi_{-n}$
where $h$ is the conformal dimension of the worldsheet field $\phi$.
In considering a product of such oscillators the order 
is not reversed.\footnote{In more general cases BPZ conjugation
is dealt with by using a reflector state representing a two
punctured disk (or sphere) with coordinates $z$ and $-1/z$ around 
$0$ and $\infty$ respectively. The conservation laws for such
state are of the type  $\langle R_{12}| (\phi_n^{(1)} - (-)^{n+h}
\phi_{-n}^{(2)} ) =0$. When having a Fock space state $|{\cal
O}\rangle_{(1)}$ 
made by the product of (possibly non-commuting) oscillators living in
state
space $(1)$, the BPZ conjugate is simply defined to be
$\langle R_{12}|{\cal O}\rangle_{(1)}$. In evaluating this state
one uses the conservation laws until all oscillators of ${\cal O}$ live
in state space $(2)$ at which point the vacuum in state space $(1)$
deletes one of the punctures leaving the out vacuum in state space
$(2)$:  $\langle R_{12}|0\rangle_{(1)} = {}_{(2)}\langle 0|$. The 
final BPZ state is a bra. Note that for non-commuting oscillators
the final ordering is the reversed one.} Thus for
example:
$bpz ( c_1|0\rangle) = 
\langle 0| c_{-1}$, and $bpz (c_0 b_{-2} c_1|0\rangle ) = 
-\langle 0| c_0 b_{2} c_{-1}$.  With these relations 
and using $\la 0|c_{-1} c_0 c_{1}|0 \ra =1$, 
the computation 
of the kinetic terms is straightforward. We find:
\be
{1\over 2\pi^2} f(T)_{kin}= -{1\over 2} 
\,  t^2  -{1\over 2} \,u^2\,
+\, {13\over 2} \,\, v^2 
+\,\, 2\, w^2\,\, +\, 3\, u\, w\, - \, 13 \, v\, w \,.
\ee

\bigskip
Let us now consider the computation of the various cubic
couplings. It will be instructive to consider the possible
use of different basis states to describe the modes of the
string field. Given the relations
\ben
c_{-1} |0\rangle  &=& {1\over 2} L_{-1}^2 \, c_1 |0\rangle\,, \cr
c_0 b_{-2} c_1 |0\rangle  
&=& \Bigl( {1\over 2} L_{-2}^{gh} - {3\over 4} L_{-1}^2
\Bigr) \, c_1 \, |0\rangle\,,
\een 
we can express all the above states in the basis
consisting of ghost and matter Virasoro descendents
of the tachyon. The computation can then be carried
out using conservation laws for matter ($c=26$) and ghost
($c=-26$) Virasoro operators. This procedure generalizes
to computations to arbitrary level, since as shown in Section 2.2
ghost and matter Virasoro descendents of the tachyon form a basis of
$\HH_{univ}^{(1)}$.

Alternatively, we can also use the ghost U(1) current $j(z) =: c(z) b(z):
= \sum
{j_n\over  z^{n+1}}$. Here $q= - 3/2$, $j_0$ is ghost number, and
$[j_n, j_m] = n\, \delta_{n+m,0}$. The tachyon state is
a U(1) primary, namely $j_{n} \, c_1 |0\rangle = 0$ for $n\geq 1$. 
A simple computation gives
\ben
\label{ghcux}
c_{-1} |0\rangle  &=& {1\over 2} \Bigl( j_{-2} + j_{-1} j_{-1} \Bigr)
 c_1|0\rangle \,, \cr\cr 
c_0 b_{-2} c_1 |0\rangle  
&=&  {1\over 2}\Bigl( j_{-2} - j_{-1} j_{-1} \Bigr)
 c_1|0\rangle \,.
\een 

Let us illustrate the various possibilities with 
the computation of the $ttu$ term in
the action. The factor of $(1/3)$ in front of the cubic
term cancels with the symmetry factor of three, which is
the number of ways the $t,t,u$ fields can be assigned
to three punctures. Given the cyclicity property the position
of $u$ is irrelevant. We therefore have to
compute 
\be
\langle c_1, c_{-1}, c_{1} \rangle 
= \langle V_3| \,\, 
c_1^{(1)} \, c_{-1}^{(2)} \,  c_{1}^{(3)}\,\,\,\,
|0\rangle_{(3)}\otimes |0\rangle_{(2)} \otimes|0\rangle_{(1)}\,, 
\ee  
where the subscripts denote the labels distinguishing the three
state spaces.
More precisely we want to relate this term to the tachyon one
given in \refb{bcor}. To this end we can simply use the conservation
law for $c_{-1}$ as given in \refb{c-1}. Note that the only
term that contributes is described by the replacement $c_{-1}^{(2)}\to
{11\over 27} c_1^{(2)}$. There are no extra signs since the
sign necessary to bring $c_{-1}$ next to the vertex bra is
compensated by the sign needed to bring the reflected oscillator $c_1$
back to the middle position. Therefore
\be
\label{jd} 
\langle c_1, c_{-1}, c_{1} \rangle = {11\over 27} \langle c_1, c_{1},
c_{1}
\rangle= {11\over 27} K^3\,. 
\ee
Therefore the corresponding term in the potential is
\be
V(T) = \cdots +{11\over 27} K^3 utt \,.
\ee 
We can repeat this calculation
describing
$c_{-1}|0\rangle$ as a $U(1)$ ghost current descendent. To this end
we note that
$q= -3/2$ and  compute
\be
\langle c_1, j_{-2}c_{1}, c_{1} \rangle = \Bigl( -{22\over 27}\,q -
{2\over
9}\,\Bigr) 
\langle c_1, c_{1}, c_{1}
\rangle=  K^3\,, 
\ee
upon use of \refb{j-2} and noticing that reflected operators into
the first and third state spaces do not contribute.
Similarly
\be
\langle c_1, j_{-1}j_{-1}c_{1}, c_{1} \rangle = - {5\over 27} 
\langle c_1, j_{1}j_{-1}c_{1}, c_{1}
\rangle=  -{5\over 27}\, K^3\,. 
\ee
The two last results, combined using the first equation in
\refb{ghcux}, agree with \refb{jd}.  Using the ghost recursion
relations one readily finds the additional terms:
\ben
\langle c_1 , c_{-1} , c_{-1}\rangle =  {19\over 243}\, K^3\,, \qquad
\langle c_{-1} , c_{-1} , c_{-1}\rangle =  {1\over 81}\, K^3\,.
\een
Incorporating such terms into the potential we now have
\be
V(T) = \cdots +\Bigl( {11\over 27}\, utt + {19\over 243}tuu + {1\over
243} uuu
\Bigr)  K^3\,.
\ee 
Let us now consider some correlators involving only matter operators.
For example, to compute the $ttv$ interaction we write
\be
\langle c_1, L_{-2}^m c_1 , c_1\rangle = \langle {\bf 1} , L_{-2}^m{\bf
1} , {\bf
1} \rangle_m \,\,\,\cdot\,\langle c_1 , c_{1} , c_{1}\rangle
\ee
since the ghost and matter correlators factorize. In the first
correlator
the normalization is $\langle {\bf 1} , {\bf 1} , {\bf
1} \rangle_m  = 1$. The matter correlator is readily computed using
the conservation of $L_{-2}$
\be
\langle {\bf 1} , L_{-2}{\bf 1} , {\bf 1} \rangle = - {5\over 54}\, c
\,,
\ee
where in the present application $c=26$. Therefore we have
\be
\langle c_1, L_{-2}^m c_1 , c_1\rangle = -{65\over 27} K^3\,.
\ee
Using the Virasoro recursion relations one readily finds
the following expressions, valid for arbitrary $c$:
\ben
\langle {\bf 1} , L_{-2}{\bf 1} , {\bf 1} \rangle &=& - {5\over 54}\, c
\equiv f_{200}(c) \,, \cr\cr
\langle L_{-2}{\bf 1} , L_{-2}{\bf 1} , {\bf 1} \rangle &=&  {128\over
729}\, c 
+ {25\over 2916}\, c^2 \equiv f_{220}(c)\,, \cr\cr
\langle L_{-2}{\bf 1} , L_{-2}{\bf 1} , L_{-2}{\bf 1} \rangle &=& 
{4096\over 19683}\, c - {320\over 6561}\, c^2 - {125\over 157464}\, c^3
\equiv
f_{222}(c)\,.
\een
These equations are easily used to produce all couplings of $v$ to
tachyons:
\ben
V(T) &=&\cdots + \Bigl(  f_{200}(26)\, vtt + f_{220}(26) \, vvt +
{1\over 3}
f_{222}(26)\, vvv \Bigr) K^3 \cr\cr 
&=&\cdots +\Bigl( -{65\over 27}\, vtt + {7553\over 729} \, vvt -
{272363\over
19683}\, vvv \Bigr) K^3  \,.
\een
As a last illustration consider couplings to the field $w$, outside the
Siegel gauge. Using the $b_{-2}$ conservation (read from that of
$L_{-2}$ with $c=0$) we have
\ben
\langle c_1, c_0b_{-2} c_1, c_{1} \rangle 
&=& -\langle V_3| \,\, 
c_1^{(1)} \, b_{-2}^{(2)}c_0^{(2)} c_1^{(2)} \,  c_{1}^{(3)}\,\,\,\,
|0\rangle_{(3)}\otimes |0\rangle_{(2)} \otimes|0\rangle_{(1)}\cr\cr
&=& + {16\over 27} \, \langle V_3| \,\, 
c_1^{(1)} \, b_{0}^{(2)}c_0^{(2)} c_1^{(2)} \,  c_{1}^{(3)}\,\,\,\,
|0\rangle_{(3)}\otimes |0\rangle_{(2)} \otimes|0\rangle_{(1)}\cr\cr
&=& + {16\over 27} \, \langle V_3| \,\, 
c_1^{(1)} \, c_1^{(2)} \,  c_{1}^{(3)}\,\,\,\,
|0\rangle_{(3)}\otimes |0\rangle_{(2)} \otimes|0\rangle_{(1)}
= {16\over 27} K^3\,.
\een
This gives the $ttw$ coefficient. The $tuw$ coefficient requires
\be
\langle   c_{-1}, c_0b_{-2}c_{1},c_1
\rangle= {16\over 81} \, K^3 \,,
\ee  
which again, is most effectively done by using first the $b_{-2}$
conservation.
Other coefficients are computed similarly.
It is perhaps worth noting that the
coefficient of $w^3$ vanishes because of the conservation $\langle V_3|
(c_0^{(1)}+ c_0^{(2)}+c_0^{(3)})=0$.

Collecting together
our results we find that the interacting part of 
the potential is given by
\ben
 {1\over 2 \pi^2 }\,f^{(4)}_{inter}(T) &=&  \Bigl(\,\,\, {1\over 3} \,
t^3
 + { 11 \over 27} \, t^2\, u
-{65 \over\,27}\,  t^2 \,v 
\,+\, {16\over 27} \, t^2\, w\cr\cr
&&\quad 
 \,\, + {19\over 243}\,t\,u^2
+ {7553\over 729}\,t\,v^2
\, + \, {64\over 243 } \,t\, w^2\cr\cr
&&\quad -{1430\over 
729} \,t\,u\,v \,\, - \,  
{2080\over 729} \, t\,v\,w \,-\, 
{32\over 81}  \,t\,u\,w\,\,  \cr\cr
&&\quad+  {1\over 243 }\,
\,u^3 \, -\,  
{272363 \over  19683 }\,\,v^3 \, +\, 0\cdot w^3 \cr\cr
&&\quad-\, {1235 \over 6561 }\, \,u^2\,v 
   +\, { 83083\over 19683}\, \,u\,v^2  
- {11248\over 19683} \, u^2\,w \, 
+\, {120848\over 19683} \, v^2 \, w \cr\cr
&&\quad -{3008\over 19683}\, u\,w^2 
- {4160\over 6561} \, v\,w^2
\, +\, {2080\over 2187} \, u\,v\,w\,\,\Bigr) \, K^3\,\,.
\een 
The string field theory is gauge invariant, but a level truncated
expression will not be so. While the gauge transformations are
nonlinear and mix all fields, to the linearized level they
take the form $\delta |\Phi\rangle = Q_B\,|\Lambda\rangle$.  With gauge
parameter
$\epsilon b_{-2}c_1|0\rangle\, $ we find
$$\delta w = \epsilon, \quad \delta u = 3\epsilon, \quad 
\delta v = \epsilon,$$
and therefore $(u-3w)$ and $(v-w)$ are gauge invariant at the
linearized level. While the Siegel gauge $w=0$ is possible, other
choices may be allowed.  In the level expansion one should not
expect all gauge choices to give the same results at each level.
Allowed gauge choices should give the same results for physical
questions in the infinite level limit. While there is no proof
that the Siegel gauge is a good gauge non-pertubatively, the
results obtained in tachyon computations certainly suggest this
is the case.

One can try to find a critical point for the above potential
without using any gauge. Had the potential been gauge invariant
one would expect undetermined parameters, but since gauge invariance
is broken by the truncation, one finds a definite critical
point\footnote{Ashoke Sen has suggested that this uncontrolled
lifting of flat directions is likely to make numerical work
based on gauge invariant actions less reliable. One could certainly
find large fluctuations in the expectation values associated with
quasi-flat
directions.}. The critical point is found at $t=0.573$, $u = -0.1074$,
$v=-0.04225$,
$w = -0.1807$, and gives 88.0\% of the required vacuum energy. In 
contrast, in the Siegel gauge, we obtain about 96\% of the
vacuum energy (with $t=0.544  , u= 0.190, 
v= 0.0560).$

\sectiono{The Identity, Star Products and Universality}

In this section we use the technology developed
before to get insight into the nature of the
identity string field.  We also learn how to
compute explicitly in the level expansion a few 
star products, including that of two zero-momentum
tachyons. We isolate a family of string fields
associated to once-punctured disks. Such surface states,
called {\it wedge states} because the local 
coordinate half-disk defines a wedge of the unit disk,
form a subalgebra of the star algebra.  
Finally, we show using conservation laws that $\HH_{univ}$ defines
a subalgebra of the star algebra.

\subsection{The so-called identity string field}

A formal object which has often been considered 
in discussions of open string field theory is the identity
element of the $*$ algebra, a string field $\Id $ 
which is formally expected to obey (with some caveats that will be 
discussed below) 
\be
\Id *  |\Phi \ra = | \Phi \ra * \Id = | \Phi \ra
\ee
for any string field $| \Phi \ra$. A Fock space expression for
 $\Id $ in terms of flat--space free--field oscillators
has been given in~\cite{gross}.
It is natural
to ask whether  $\Id $ can be given a manifestly
background--independent representation. 

The answer is immediate  when we  recall
the geometric interpretation of the identity. 
In the  Schr\"{o}edinger representation, 
the functional $\Psi_{\II}[\phi(\sigma)]$ associated to the identity 
acts on an open--string configuration $\phi(\sigma)$ ($\phi$ is
a collective name for the matter and ghost fields) as a delta
function overlap of the left and right halves of the string.
In other terms,  $\Psi_{\II}[\phi(\sigma)]$ vanishes unless
the left ($0 \leq \sigma \leq \pi/2$)
and right  ($\pi/2 \leq \sigma \leq \pi$)
halves of the string exactly coincide. In
CFT language, such functional is represented by a state
$\la \II| \in\HH^*$ satisfying 
$ \la \II | \, \Phi \ra \equiv \la  \Phi \ra $ for all $|\Phi \ra \in \HH$,
where the one point function is computed on a specific 1--punctured
disk, one where the local coordinates are such that the left and
right halves of the worldsheet boundary are glued together.
$\la \II |$ is a surface state, since just as $\la V_3|$ it
encodes the correlators on a particular Riemann surface.    

Representing as usual the worldsheet 
as the upper half-disk $\{ |z| \leq 1, \Im z \geq 0 \}$,
the function
\be
w= F^{360^\circ}(z) =\left( \frac{1+iz}{1-iz} \right)^2
\ee
sends this upper--half unit disk to the full unit disk
in the $w$ plane, 
mapping the  two halves of the string ($\{ |z| =1, \, \Re z > 0 \}$ 
and $\{ |z| =1,\, \Re z < 0 \}$) to the same interval $\{ \Im w= 0, -1
\leq \Re w \leq 0 \}$. It is convenient to map
back the disk to the upper half--plane. 
Our final choice of local coordinate
is then
\be \label{f360}
\tilde z =  f^{360^\circ}(z) = h^{-1}(F^{360^\circ}(z)) =
\frac{2 z}{1-z^2} \,,
\ee
where $h$ was defined in \refb{hdefi}.
The puncture is at $\tilde z =0$, and the image of 
the unit upper half disk is the full upper
half $\tilde z$-plane. 

An explicit representation 
for the identity is now easy to write. 
If we can find the operator
$U_{  f^{360^\circ}}$ that implements the
conformal transformation $f^{360^\circ}(z)$ in the CFT state space,
we have
\be
\la \II | = \la 0| \,U_{  f^{360^\circ}} \,.
\ee
Such operator must be written as the exponential of  
a linear combination
of the total (matter + ghost) Virasoro generators \cite{LPP}  
\be
U_{  f } =
e^{v_0 L_0} e^{\sum_{n \geq 1} v_n L_n} \,.
\ee
This makes manifest the background independence of $\la \II |$.
Since $f(0)=0$ only
positively moded Virasoro generators enter $U_f$,  and we have
chosen to separate
out the global scaling component $e^{v_0 L_0}$.
The coefficients $v_n$ can be determined recursively
from the Taylor expansion of $f$,         
by requiring \cite{LPP, sen2}
\ben
e^{v_0}& =& f'(0) \nonumber \\
\exp\left({\sum_{n \geq 1} v_n z^{n+1} \partial_z} \right)\, z & =&  
[f'(0)]^{-1} f(z) = z+ a_2 z^2 + a_3 z^3 + \dots\,\,.
\een
For example, for the first coefficients one finds
\be
v_1=a_2 \,, \quad
v_2=-a_2^2+a_3 \,, \quad 
v_3 = \frac{3}{2}\, a_2^3 
 -\frac{5}{2}\, a_2 a_3 +a_4 \,\,.
\ee
Taking $f=f^{360^\circ}$ we obtain
\be
U_{f^{360^\circ}} = 2^{L_0} \exp
\Bigl( L_{2} - {1\over 2} L_{4} + {1\over 2} L_{6} - {7\over 12}
L_{8}
+ {2\over 3} L_{10} + \cdots \Bigr) \,.
\ee
By SL(2,R) invariance of the vacuum, $\la 0| L_0 =0$,
so that $\la \II |$ does not in fact
depend on the overall scaling factor $v_0$.\footnote{Incidentally,  
SL(2,R) invariance also guarantees that  
the surface state is independent of the
particular SL(2,R) frame used to write the map $f$. If
$\tilde f= R \circ f$, where $R$ is an SL(2,R) transformation,
then $\la 0| U_{\tilde f} = \la 0 | U_{R} U_{f} =\la 0| U_{ f}$.
The composition law 
$U_{g \circ f}=U_g U_f$ holds in a CFT with vanishing central charge,
as is the case here.}

To obtain the ket $\Id$ simply recall that BPZ conjugation
sends $L_n$ to $(-1)^n L_{-n}$,
\be \label{identity}
\Id = 
 \exp
\Bigl( L_{-2} - {1\over 2} L_{-4} + {1\over 2} L_{-6} - {7\over 12}
L_{-8}
+ {2\over 3} L_{-10} + \cdots \Bigr) |0 \ra \,.
\ee
This expression is  
well--defined in the level truncation scheme. Only a finite
number of terms in the sum $\sum_n v_n L_{-n}$ are needed to write
the expression of $\Id$ truncated at some given level\footnote{This
property is not immediately obvious for the Fock space 
expression of \cite{gross}.}. 

The formalism of conservation laws allows to deduce
several properties of the identity. Let us begin with
Virasoro conservation laws. Let $\tilde z$ be as in (\ref{f360})
the global coordinate on the 1--punctured disk (upper half plane)
associated to $\la \II |$, and $z$ the local coordinate around
the puncture. For any vector field $\tilde v(\tilde z)$,
holomorphic everywhere (including infinity), except for a possible
pole at the puncture $\tilde z =0$
we have 
\be
\la \II | \, \oint_{{\cal C}} \tilde T (\tilde z) \tilde v (\tilde z)  
d \tilde z =0\,,
\ee
where ${\cal C}$ is a contour circling the origin. As usual, this
statement
is obtained  by shrinking the contour to the point at infinity.
By passing to the local coordinate $z$ we deduce
\be \label{basicidentity}
\la  \II    | \; \oint_{{\cal C}}   dz \; v(z)  
\, \Bigl( T(z) - {c\over 12} \,
S(f^{360^\circ}(z),z)\Bigr)   \; = 0 \;.
\ee
We have $S(f^{360^\circ}(z), z)= 6 (z^2 +1)^{-2} = 6(1-2z^2+3z^4-4z^6
+\dots)$.
Taking $\tilde v(\tilde z) = \tilde z$, $\tilde v(\tilde z) = 2$,
and $\tilde v(\tilde z) = 4/\tilde z$, respectively, we
find
\ben
\la \II | \, \left(L_{0} -2 L_2 + 2 L_4 - 2 L_6 + \dots \right)\; &=&
0
\;,\cr
\la \II | \, \left(L_{-1} -3 L_1 + 4 L_3 - 4 L_5 + \dots \right)\;
&=&  0
\;,\cr
\la \II | \, \left(L_{-2} -(c/2) -4 L_0 + 7 L_2 - 8 L_4 + \dots
\right) \; &=& 0 \;.
\een
One can use (\ref{identity}) to check
these conservation laws in the level expansion. 
In addition, for $c=0$ Virasoro generators we
also have
\be
\label{konid}
\la \II | K_n = 0\,.
\ee
This is proved as follows. We consider the vector
field $v(z) = z^{n+1} - (-)^n z^{-n+1}$ as before
and confirm that: (i) it satisfies $v(-1/z) z^2 = v(z)$ as
required by the identification $z\to -1/z$, (ii) the vector
field has no poles anywhere else in the $\tilde z$ plane.
Checking (ii) requires verifying this is the case both
for $\tilde z = i$ (in fact, $v(\tilde z =i)$ is zero))
and for $\tilde z = \infty$ (in fact for $\tilde z \to \infty$,
$v(\tilde z) \sim \tilde z^2$, which is regular
at $\tilde z \to \infty$).

Let us now consider the action of ghost
oscillators on the identity. While the above argumentation
suggests conservation laws based on the $C_n$ operators
(see (\ref{Cndef}, \ref{Cn})) the quadratic differential 
(\ref{Cndiff}) would
actually have poles at $\tilde z = i$ invalidating the 
conservation. We therefore start from basics, the
conservations take the form
\be
\label{conI}
\la  \II    | \; \oint_{{\cal C}}   dz \; \phi(z) c(z) \;=0 \;,
\ee
where $\phi(z)$ is a quadratic differential holomorphic everywhere 
except at the puncture, and $z$ is again the local
coordinate defined in (\ref{f360}). Consider quadratic
differentials
of the form $\tilde \phi
(\tilde z) = 1/\tilde z^n$, manifestly regular at  $\tilde z = i$.
Since the quadratic differential must not have a pole at $\tilde z =
\infty$ we must require
$n\geq 4$. Let us consider the simplest one:
\be \tilde \phi
(\tilde z) = 1/\tilde z^4 \quad \to \quad \phi(z) = {(1+ z^2)^2\over
2 z^4}.
\ee
Note that this quadratic differential is well-defined in the
glued surface, as $\varphi = \phi(z) dz^2$ is invariant under
the gluing that identifies $z$ with $-1/z$
on the halves of $|z| = 1, \Im z >0$.  Back in \refb{conI} we
get
\be 
\label{conv}
\la  \II    | \;\Bigl( c_0 + {1\over 2} (c_2 + c_{-2} ) \Bigr) \;=0
\;.
\ee
It follows from this conservation law that
$c_0 | \II \ra \not= 0$. This, in fact, shows that $|\II \ra$ {\it is
not an identity of the star algebra on all states}.
Indeed,
consider the value of $c_0 (\II * A)$, which must equal $c_0 A$ if
$\II$ is an identity for $A$.
Using \refb{c0}, however, we actually find that $c_0 (\II * A) = 
c_0\II * A + \II* c_0 A$. If $\II$ is also an identity for $c_0 A$,
it follows that  $c_0 \II * A = 0$ for all $A$. While this might
actually be
true for Fock space states $A$,\footnote{We thank M. Schnabl for 
raising this possibility.} for $A= \II$ it implies that $c_0\II$
vanishes, which is patently false. Such failure of formal
properties is certainly familiar in open string field theory \cite{failure}.

Finally, we  consider a holomorphic current $\tilde \jmath(\tilde z)$ 
and a holomorphic scalar $\tilde f(\tilde z)$  
and find
\be 
2 q\, \la  \II    | \; \oint_{\tilde w =0}  d\tilde w \; {\widehat f(\tilde
w)\over \tilde w}+ \la  \II    | \; \oint_{{\cal
C}}   dz \; f(z) \left( j(z) -q \, 
\frac{d^2 f^{360^\circ} }{ dz^2  } \left( \frac{d f^{360^\circ} }{ dz
}\right)^{-1} \right) \;=0 \;.
\ee
Here $z$ is the local coordinate, $\tilde z= f^{360^\circ}(z)=2 z/(1-z^2)$
and $\tilde w = 1/\tilde z$.
The functions $\widehat f$ and $f$ are conformal tranforms of $\tilde f$ 
defined by the scalar tranformation laws 
$\widehat f(\tilde w) =\tilde f(\tilde z)$ and $f(z) = \tilde f(\tilde z)$.
 The simplest conservation arises for a constant $\tilde f$, 
giving
$\la \II | ( 2q + j_0 ) = 0$. For  the BRST current
$j_B(z)$, we have $ j^0_B=Q_B$, and $q=0$. This gives
\be
\la  \II    | \; Q_B =0\,.
\ee 
This property
is also manifest from the representation (\ref{identity}),
since $[Q_B, L_n ] =0$ and $Q_B |0 \ra =0$.

\subsection{A subalgebra of wedge states}

The observation that
the identity is the surface
state associated with the map $F^{360^\circ}$ 
naturally leads us  to consider a generalization to `wedge--like'
states of arbitrary angle. As we shall explain, 
this family of states, arising from once punctured
disks,  has the
interesting property of being a subalgebra of the $*$-algebra.

We begin by considering the map
\be
w= F^{\frac{360^\circ}{n}}(z) =
\left( \frac{1+iz}{1-iz} \right)^{2 \over n}\,,
\ee 
which sends the unit upper half disk in the $z$ plane to 
a wedge of angle $\frac{360^\circ}{n}$ in the $w$ plane. 
Such maps are used in the higher point interactions of 
open superstring field theory \cite{berkovitssupersft}. As usual, 
we map back to the upper half plane:
\be \label{f360n}
\tilde z =  f^{{360^\circ \over n}}(z) = h^{-1}(F^{{360^\circ \over
n}}(z)) =
\tan \left( {2 \over n} \arctan(z) \right) \,.
\ee
Let us  then define the family of surface states
\be
\Bigl \la \frac{360^\circ}{n} \Bigr | \equiv   \la 0| U_{
f^{\frac{360^\circ}{n}}} \,.
\ee
With the same method used to calculate the
identity we can write an explicit expression for these wedge states:
\be
\label{nwedge}
\Bigl|{ 360^\circ \over n} \Bigr\rangle = \exp
\Bigl( -{n^2-4\over 3n^2}\, L_{-2} + {n^4 -16\over 30n^4}\, L_{-4} -
{(n^2-4) (176+ 128 n^2 + 11 n^4)\over 1890 n^6} \,L_{-6}
 + \cdots \Bigr) | 0\rangle 
\ee
For $n=1$ we recover the identity: $|{360^\circ}\rangle = |\II\ra$
(see \refb{identity}).
For $n=2$ we get the vacuum:  $|180^\circ\rangle = |0\rangle$.
For $n \to \infty$ we find a smooth limit
\be
|\frac{ 360^\circ }{\infty} \rangle = \exp
\Bigl( -{1\over 3} L_{-2} + {1\over 30} L_{-4} - {11\over 1890} L_{-6} +
{1\over 1260} L_{-8} + 
{34\over 467775} L_{-10}
 + \cdots \Bigr) | 0\rangle \,.
\ee
The existence of the $n \to \infty$ limit can be understood
from the expression for the conformal map,
\be
 f^{{360^\circ \over n}}(z) \stackrel{n \to \infty}{\longrightarrow} \frac{2}{n}
\arctan(z) \,.
\ee
The map has a well--defined limit up to a vanishing scaling
factor, which is immaterial in the definition of the surface state.

The wedge surface states form a subalgebra 
of $\HH_{univ}^{(0)}$, in particular we claim
 that\footnote{Here $r$ and $s$
will denote real numbers larger than or equal to one.} 
\be
\label{wsa}
\Bigl|{360^\circ \over r_1} \Bigr\rangle * \Bigl|{360^\circ \over r_2}
\Bigr\rangle = \Bigl|{360^\circ\over {r_1+r_2-1}} \Bigr\rangle\,.
\ee
This is readily understood by sewing the appropriate surfaces.
To sew wedges, however, we must first use conformal transformations
of power type to eliminate corners. Let the input states above
be defined on  disks $\xi_i$ with local coordinates $\eta_i$: 
$\xi_i(\eta_i) = F^{360^\circ\over r_i}(\eta_i) $ ($i=1,2$). We must then
introduce new disks $w_i = (\xi_i)^{r_i/2}$. It then follows that 
$w_i(\eta_i) = F^{180^\circ}(\eta_i)$, and while the total neighborhood
angle at $w_i=0$ is $\pi r_i$, the image of the local $\eta_i$-half disk
is a $180^\circ$ wedge in $w_i$. These are now ready for sewing.  On
the three string vertex with disk $w$ we introduce a new disk
$W = w^{3/2}$ where each of the $120^\circ$ wedges, the images of
the local coordinates $z_i$, grows to a $180^\circ$ wedge. The gluing
necessary to take the product is simply $z_i \eta_i = -1 (i=1,2)$.
This is simply implemented by amputating the two $180^\circ$ wedges
on the vertex $W$, amputating the two $180^\circ$ wedges on $w_i$ and 
gluing the left-overs of the $w_i$ disks to the leftover of the $W$
disk. The glued surface has a total neighborhood angle of
$\pi(r_1-1) + \pi(r_2-1) + \pi = \pi(r_1+ r_2 -1)$, with the local
coordinate giving the last contribution of $\pi$.  Mapping back to a $360^\circ$
disk, we see that this is simply the wedge state
$\Bigl|{360^\circ\over {r_1+r_2-1}} \Bigr\rangle$ .

More generally, we could consider the family of arbitrary
surface states of the CFT, {\it i.e.}, states of the form 
$\la f | \equiv \la 0| U_f $ for a generic conformal map $f$. The gluing theorem
\cite{LPP}
guarantees that this family of string fields forms a closed subalgebra.
The product $|f \ra * |g \ra$ is equal to another surface state
$| h \ra$, where the conformal map $h$ is implicitly determined from $f$
and $g$ by the gluing procedure. It would be
interesting to find a more explicit characterization of $h$.

\subsection{Star products and $\HH_{univ}$ }

The wedge state $\Bigl|{360^\circ \over 3}\ra = |120^\circ\ra$ has an
interesting interpretation. Consider the $*$-product of two vacuum
states. 
We immediately have
\be
|0\ra * |0\ra = |180^\circ\ra * |180^\circ\ra = |120^\circ\ra\,,
\ee
where we made use of \refb{wsa}. 
This answer is easily understood if we recall that the SL(2,R) vacuum
deletes punctures.
It  then follows from the disk presentation of the vertex
(Fig.\ref{3wedgesfig})
that the resulting surface is a once punctured disk 
with a $120^\circ$ wedge.
Using the $n=3$ case of \refb{nwedge} we get:
\be \label{0*0}
 |0 \ra * |0\ra  =   \exp
\left(               -\frac{5}{27} L_{-2} 
+\frac{13}{486}L_{-4} -
\frac{317}{39366}L_{-6} + \frac{715}{236196}L_{-8}
+\dots \right)      |0 \ra \,.
\ee

At this time we can compute a couple of
additional star products. Consider the evaluation
of $|0\rangle * c_1 |0\rangle$. The idea is to
use a ghost conservation law to remove the $c_1$ from
the state space where it appears and take it into
the state space corresponding to the output. More precisely, 
in the expression
\be
|0\rangle * c_1 |0\rangle = \la V_{123'} | \Bigl( | 0\ra_{(1)}
\,\otimes \, |R_{3'3}\ra \otimes c_1^{(2)} |0\ra_{(2)} \Bigr) 
\ee
we need a conservation law where the $c_1$ can be traded
for oscillators that annihilate the vacuum in the first
state space. Such conservation arises from the quadratic
differential $\varphi (z) = {9\sqrt{3}\over 2}  \, z^{-1}
(z+\sqrt{3})^{-3}$:
\ben
&& 0=\la V_3 | \Bigl( {4\over 3\sqrt{3} }\,\,c_2 + \cdots \Bigr)^{(1)}
 +\la V_3 |\Bigl( c_{1} + \cdots \Bigr)^{(2)}\cr\cr
&&\quad-\la V_3 |\Bigl(  {27\over 16}\,\,c_{-1} + {3\sqrt{3}\over 8}\,
c_0 - {11\over 16}\, c_1 - {2\over 3\sqrt{3}}\, c_2 +  {5\over
9}\, c_3 + {44\over 81 \sqrt{3}}\, c_4 + \cdots \Bigr)^{(3)}
\een
Recalling that acting on the reflector $|R_{33'}\ra$, we have
$c_n^{(3)}
\to c_{-n}^{(3')}(-)^{n+1}$ we obtain: 
\be
|0\rangle * c_1 |0\rangle = \Bigl( {27\over 16}\,\,c_{1} -
{3\sqrt{3}\over 8}\, c_0 - {11\over 16}\, c_{-1} + {2\over
3\sqrt{3}}\, c_{-2} +  {5\over 9}\, c_{-3}  - {44\over 81 \sqrt{3}}\,
c_{-4} \cdots \,\Bigr) 
(|0\rangle * |0\rangle) 
\ee
Note that this product manifestly lies on $\HH_{univ}$. 
Using the quadratic differential  $\varphi (z) = 9\sqrt{3} \, (z-\sqrt{3})^{-1}
(z+\sqrt{3})^{-3}$ we find
\ben
&& 0=\la V_3 |\Bigl( c_1 + \cdots \Bigr)^{(1)}
 +\la V_3 |\Bigl( - {4\over 3\sqrt{3}} c_2 + \cdots \Bigr)^{(2)}\cr\cr
&&\quad-\la V_3 |\Bigl(  {27\over 16}\,\,c_{-1} - {3\sqrt{3}\over 8}\,
c_0 - {11\over 16}\, c_1 + {2\over 3\sqrt{3}}\, c_2 +  {5\over
9}\, c_3 - {44\over 81 \sqrt{3}}\, c_4 + \cdots \Bigr)^{(3)}
\een
This conservation law enables us to write 
\be
c_1|0\rangle * c_1 |0\rangle = \Bigl({27\over 16}\,\,c_{1} +
{3\sqrt{3}\over 8}\, c_0 - {11\over 16}\, c_{-1} - {2\over
3\sqrt{3}}\, c_{-2} +  {5\over 9}\, c_{-3} + {44\over 81 \sqrt{3}}\,
c_{-4}+ \cdot\cdot 
\Bigr) \,  (|0\rangle * c_1|0\rangle) \,.
\ee
Combining the results above we can write
\begin{eqnarray}
c_1|0\rangle * c_1 |0\rangle\hskip-15pt  
&&= \,{81\sqrt{3}\over 64}\, \Bigl(  c_0 
- {16\over 27}\, c_{-2}  + {352\over 729}\, c_{-4} + \cdots\Bigr) \cr\cr
&&\qquad\qquad  \Bigl( c_{1}  -
{11\over 27}\, c_{-1} +  {80\over 243}\, c_{-3}  + \cdots \Bigr)  \bigl(|0\rangle
* |0\rangle\bigr) \,.
\end{eqnarray}
This is a formula for the product of two zero momentum
tachyons. One can recognize the two factors acting
on $|0\ra * |0\ra$ as the factors that arise in the 
$c_{-1}$ and $c_0$ conservations at the special punctures
(see  (\ref{firstc}) and (\ref{c-1})).

The geometric interpretation of $|0\ra * | 0 \ra$
can also be used to find an alternative
expression for  $c_1 |0 \ra * c_1 | 0 \ra$.
This product is a $120^\circ$ wedge, with puncture at $w=1$, and two
$c$'s inserted at the other two punctures. This gives
\be
bpz(c_1 |0 \ra * c_1 |0 \ra) =
\la 0 | c(e^{\frac{2 \pi i}{3}}) c(e^{-\frac{2 \pi i}{3}})
U_{F^{120^\circ}} 
\left( \frac{d F_1^{120^\circ}(0)}{dz}  \frac{d F_3^{120^\circ}(0)}{dz}
\right)^{-1} \, ,
\ee
or equivalently, using the upper--half plane representation of the vertex
\be
bpz(c_1 |0 \ra * c_1 |0 \ra) =
\la 0 | c(\sqrt{3}) c(-\sqrt{3})
U_{f^{120^\circ}} 
\left( \frac{d f_1^{120^\circ}(0)}{dz}  \frac{d f_3^{120^\circ}(0)}{dz}
\right)^{-1} \, \,.
\ee
The derivative factors arise from the conformal transformations
of $c_1 |0 \ra$, which is a primary field of dimension $-1$.
We find
\ben 
\label{c1*c1}
 c_1 | 0 \ra * c_1 | 0 \ra & =&  
\frac{9}{64}\cdot \exp \left( v^{120^\circ}_n L_{-n} \right)
  \left( \frac{2}{3 } \right)^{L_0} 3^2 c(\frac{1}{\sqrt{3}}) 
c(-\frac{1}{\sqrt{3}}) |0 \ra 
\nonumber \\
 &&  = \frac{27}{32}\cdot\exp \left(-\frac{5}{27} L_{-2} 
+\frac{13}{486}L_{-4} -\frac{317}{39366}L_{-6} + 
\frac{715}{236196}L_{-8}+\dots \right)\cdot  \nonumber \\
&& \quad \cdot \sum_{i=0}^{\infty} \sum_{j=0}^{\infty} 
\left( \frac{2}{3\sqrt{3}} \right)^{2i+2j-1}  c_{-2j}c_{-2i+1}  |0 \ra
\,.
\een
Unlike the case of wedge states, the scaling component
$\left( \frac{2}{3} \right)^{L_0}$ of $U_{f^{120^\circ}}$
cannot be ignored.

\bigskip
By thinking of the evaluation of star products using
the conservation laws one readily sees that 
$\HH_{univ}$ is a
subalgebra,
$\HH_{univ}* \HH_{univ} \subseteq
\HH_{univ}$. Indeed, any state  in $\HH_{univ}$ 
can be obtained (by definition) acting on the vacuum with $b_{-k},
L^m_{-k}$
($k\geq 2$) oscillators and $c_{-l}$ ($l\geq -1$) oscillators.
To compute the
product 
$|\Psi_1 \ra * |\Psi_2 \ra$, with  $\Psi_i \in \HH_{univ}$, we can
use antighost and Virasoro conservation laws to move all 
the $b$, $L^m$ oscillators 
out of the state--spaces  $(1)$ and $(2)$ (where 
$|\Psi_1 \ra$ and $|\Psi_2 \ra$ are inserted)
onto the third state space. This is possible since we have 
conservation laws for $b_{-k}$ and 
$L^m_{-k}$, when $k\geq 2$. For the ghost oscillators
we can use the conservation laws for $c_0, c_{-1}, c_{-2} , \cdots$
to leave only $c_1$ oscillators possibly acting on spaces
$(1)$ and $(2)$. Thus any product can be reduced to
$c,b,L$ oscillators acting on either $c_1|0\ra * c_1 |0\ra$,
$|0\ra * c_1 |0\ra$, $c_1 |0\ra * |0\ra$ or $|0\ra * |0\ra$.
Since we have seen that any of those states are indeed in
$\HH_{univ}$ they will remain in $\HH_{univ}$ after action
by $b,c$ and $L^m$ oscillators. This proves
$\HH_{univ}$ is a subalgebra. 

Since ghost number is additive under star multiplication 
($gh(A*B) = gh(A) +
gh(B)$) the subspace $\HH_{univ}^{(0)}$ of ghost number zero states in
$\HH_{univ}$ is itself a subalgebra. 
There is an even smaller universal subalgebra at ghost number
zero. Consider
\ben
 \HH^{(0)}(L) \equiv {\rm Span} \{ L^{tot}_{-j_1} \dots L^{tot}_{-j_p}\,
 | 0 \ra\, ,  \; j_i \geq 2 \} \,.
\een  
Here $L^{tot}$ denotes the combined matter and ghost ($c=0$)
Virasoro operators. Indeed since $|0\ra *|0\ra\in \HH^{(0)}(L)$ (see
\refb{0*0}) it follows from the conservation laws that any 
product of descendents of the vacuum will be a descendent
of $|0\ra *|0\ra\in \HH^{(0)}(L)$ and thus a descendent of the vacuum.
This confirms $\HH^{(0)}(L)$ is a subalgebra. It would be
interesting to investigate it concretely. Note that neither matter
nor ghost Virasoro descendents of the vacuum form subalgebras. 

\bigskip

\section{Concluding Remarks}

In this paper we have developed a
computational scheme for string field theory.
We hope this method will be used by
physicists interested in string field theory but
previously mystified by the technical complexities of the requisite
computations. We believe that these conservation
laws are both easy to use for low level by-hand calculations
and will be straightforward to implement for
high level computer calculations. Such high
level computations should be relevant in the
near future as we are learning how to use string field
theory for non-perturbative computations.

Conservation laws can also be used for
string field computations in explicit backgrounds.
For example, the current conservation identities described
in section 4.2 can be used to compute with oscillators
associated to free bosons $i\partial X$. It would be
of interest to know if $\la V_3|$, when restricted to $\HH_{univ}$, 
can be written  as some sort of exponential of
Virasoro operators.

While we have not developed the details here, our
methods should be applicable to computations in 
superstring field theory. The relevant string field theory
is nonpolynomial \cite{berkovitssupersft}, but 
since no antighost nor picture
changing operators need to be inserted on the
world sheet, the conservation laws discussed in
this paper apply with minor modifications. It would
certainly be desirable to test the brane anti-brane 
annihilation conjecture beyond the present accuracy
of about 90\% \cite{0001084,0002211,0003220,0004015}.  

The present paper was motivated by a desire to
find an analytic expression for the tachyon condensate
in string field theory. Since the solution is an
element of the universal subalgebra $\HH_{univ}$ we
were led to believe that finding such solution 
would require a computational scheme that used
$\HH_{univ}$ and not a background dependent representation.
We have developed this computational scheme, and
have used it to learn about the identity element, to
compute some explicit star products and to identify
a subalgebra of $\HH_{univ}$. Further developments
may be needed to be able to find the exact 
tachyon condensate. Such condensate would 
represent  the first nontrivial
analytic solution of string field theory.

\bigskip 

\bigskip

\noindent {\bf Acknowledgments}:
We are indebted to Ashoke Sen for many
conversations and detailed discussions on the subject
presented in this paper. We also acknowledge 
instructive conversations with M. Douglas, A. Iqbal, N. Moeller and
W. Taylor.~~The work of B.Z. and L.R. was supported in part
by DOE contract \#DE-FC02-94ER40818.


\begin{thebibliography}{99}


\bibitem{9902105}
A.~Sen,
``Descent relations among bosonic D-branes,''
Int.\ J.\ Mod.\ Phys.\  {\bf A14}, 4061 (1999)
[hep-th/9902105].


\bibitem{RECK}
A.~Recknagel and V.~Schomerus,
``Boundary deformation theory and moduli spaces of D-branes,''
Nucl.\ Phys.\ {\bf B545}, 233 (1999)
hep-th/9811237; \\
C.G.~Callan, I.R.~Klebanov, A.W.~Ludwig and J.M.~Maldacena,
``Exact solution of a boundary conformal field theory,''
Nucl.\ Phys.\ {\bf B422}, 417 (1994)
hep-th/9402113; \\
J.~Polchinski and L.~Thorlacius,
``Free fermion representation of a boundary conformal field theory,''
Phys.\ Rev.\ {\bf D50}, 622 (1994)
hep-th/9404008.



\bibitem{9805019}
A.~Sen,
``Stable non-BPS bound states of BPS D-branes,''
JHEP {\bf 9808}, 010 (1998)
[hep-th/9805019].


\bibitem{9805170}
A.~Sen,
``Tachyon condensation on the brane antibrane system,''
JHEP {\bf 9808}, 012 (1998)
[hep-th/9805170].


\bibitem{9808141}
A.~Sen,
``SO(32) spinors of type I and other solitons on brane-antibrane
pair,''
JHEP {\bf 9809}, 023 (1998)
[hep-th/9808141].

\bibitem{9810188}
E.~Witten,
``D-branes and K-theory,''
JHEP {\bf 12}, 019 (1998)
hep-th/9810188.

\bibitem{9812031}
A.~Sen,
``BPS D-branes on non-supersymmetric cycles,''
JHEP {\bf 12}, 021 (1998)
hep-th/9812031.

\bibitem{9812135}
P.~Horava,
``Type IIA D-branes, K-theory, and matrix theory,''
Adv.\ Theor.\ Math.\ Phys.\ {\bf 2}, 1373 (1999)
hep-th/9812135.

\bibitem{9912249}
A. Sen and B. Zwiebach, 
``Tachyon Condensation in String Field Theory,'' hep-th/9912249.

\bibitem{0001201}
W. Taylor, ``D-brane effective field theory from string field theory", 
hep-th/0001201. 


\bibitem{0002237}
N.\ Moeller and W.\ Taylor, {
``Level truncation and the tachyon in open bosonic string field
theory"},
hep-th/0002237.

\bibitem{0002117}
J.A.\ Harvey and P.\ Kraus, {
``D-Branes as unstable lumps in bosonic open string field theory"},
hep-th/0002117.

\bibitem{0003031}
R.~de Mello\ Koch, A.\ Jevicki, M.\ Mihailescu and R.\ Tatar,
{``Lumps and p-branes in open string field theory"},
hep-th/0003031.

\bibitem{0001084}
N. Berkovits, ``The Tachyon Potential in Open
Neveu-Schwarz String Field Theory,'' [hep-th/0001084].

\bibitem{0002211}   
N.\ Berkovits, A.\ Sen and B.\ Zwiebach, {
``Tachyon condensation in superstring field theory"},
hep-th/0002211.

\bibitem{0003220}    
P.~De Smet and J.~Raeymaekers,
{``Level four approximation to the tachyon potential in superstring
field theory"},
hep-th/0003220.

\bibitem{0004015}
A.~Iqbal and A.~Naqvi,
``Tachyon condensation on a non-BPS D-brane,''
hep-th/0004015.

\bibitem{0005036}
N.~Moeller, A.~Sen and B.~Zwiebach,
hep-th/0005036.

\bibitem{chicagoBfield}
J.~A.~Harvey, P.~Kraus, F.~Larsen and E.~J.~Martinec,
``D-branes and strings as non-commutative solitons,''
hep-th/0005031.


\bibitem{wittenBfield}  
E.~Witten,
``Noncommutative tachyons and string field theory,''
hep-th/0006071.


\bibitem{KS}
V.A. Kostelecky and S. Samuel, ``The Static Tachyon Potential in
the Open Bosonic String Theory,'' Phys. Lett. {\bf B207} (1988) 169; \\
V.~A.~Kostelecky and R.~Potting,
``Expectation Values, Lorentz Invariance, and CPT in the Open Bosonic
String,''
Phys.\ Lett.\  {\bf B381}, 89 (1996)
[hep-th/9605088].

\bibitem{WITTENBSFT}
E.~Witten,
``Noncommutative Geometry And String Field Theory,''
Nucl.\ Phys.\  {\bf B268}, 253 (1986).

\bibitem{gross}
D. Gross and A. Jevicki, ``Operator formulation of interacting
string field theory (I), (II)," Nucl.\ Phys.\ {\bf B283} (1987) 1, 
{\bf B287} (1987) 225.

\bibitem{cremmer}E. Cremmer, A. Schwimmer and C. Thorn, 
``The vertex function
in Witten's formulation of string field theory," 
Phys. Lett. {\bf B179} (1986) 57.

\bibitem{samuel}
S. Samuel, ``The physical and ghost vertices in Witten's string 
field theory,", Phys. Lett. {\bf B181} (1986) 255.

\bibitem{LPP}
A.~LeClair, M.E.~Peskin and C.R.~Preitschopf,
``String Field Theory On The Conformal Plane. 1. Kinematical
Principles,''
Nucl.\ Phys.\ {\bf B317}, 411 (1989); 
``String Field Theory On The Conformal Plane. 2. Generalized Gluing,''
Nucl.\ Phys.\ {\bf B317}, 464 (1989).

\bibitem{9911116}
A. Sen, ``Universality of the Tachyon
Potential,'' [hep-th/9911116].


\bibitem{AlvarezGaume:1988bg}
L.~Alvarez Gaume, C.~Gomez, G.~Moore and C.~Vafa,
``Strings In The Operator Formalism,''
Nucl.\ Phys.\  {\bf B303} (1988) 455.

\bibitem{Vafa:1987ea}   
C.~Vafa,
``Conformal Theories And Punctured Surfaces,''
Phys.\ Lett.\  {\bf B199} (1987) 195.


\bibitem{strebel}
K.~Strebel, {\it Quadratic differentials}, Springer Verlag (1984).


\bibitem{gmw}
S. Giddings, E. Martinec and E. Witten, ``Modular invariance
in string field theory'', Phys. Lett. {\bf B176} (1986) 362.


\bibitem{Zwiebach:1991az}
B.~Zwiebach,
``A proof that Witten's open string theory gives a single cover of
moduli space,'' Commun.\ Math.\ Phys.\  {\bf 142}, 193 (1991).



\bibitem{gsw} M. Green, J. H. Schwarz, and E. Witten, ``Superstring
Theory'', Cambridge University Press.


\bibitem{wittensupersft}
E.~Witten,
``Interacting Field Theory Of Open Superstrings,''
Nucl.\ Phys.\  {\bf B276}, 291 (1986).



\bibitem{sen2} A. Sen, ``Open String Field Theory in a non--trivial
Background Field, (I) '', Nucl. Phys. {\bf B334} (1990) No.2, 350. 



\bibitem{failure}
G.~T.~Horowitz and A.~Strominger,
``Translations As Inner Derivations And 
Associativity Anomalies In Open String
Field Theory,''
Phys.\ Lett.\  {\bf B185}, 45 (1987).


\bibitem{berkovitssupersft}
N. Berkovits, ``Super-Poincare Invariant Superstring Field Theory,''
Nucl. Phys. {\bf B450} (1995) 90, [hep-th/9503099].


N. Berkovits,
``A New Approach to Superstring Field Theory,''
proceedings to the
$32^{nd}$ International Symposium Ahrenshoop on the
Theory of Elementary Particles, Fortschritte der Physik (Progress of
Physics) {\bf 48} (2000) 31, [hep-th/9912121].


N. Berkovits and C.T. Echevarria, 
``Four-Point Amplitude from Open Superstring
Field Theory,'' [hep-th/9912120].





\end{thebibliography}
\end{document}